\documentclass[aps,prl,10pt,twocolumn,titlepage,showkeys,nofootinbib,showpacs,superscriptaddress]{revtex4-1}
\usepackage[left=1.7cm,right=1.7cm,top=1.2cm,bottom=1.2cm,includeheadfoot]{geometry}
\usepackage{graphicx}
\usepackage{url}
\usepackage{amsmath}
\usepackage{amssymb}
\usepackage{bm}
\usepackage{dsfont}
\usepackage{dcolumn}
\usepackage{calc}
\usepackage[pdftex,table]{xcolor}
\usepackage{array}
\usepackage{booktabs}
\usepackage{courier}
\usepackage{scalefnt}
\usepackage[titletoc]{appendix}
\newcommand*\phantomas[3][c]{\ifmmode\makebox[\widthof{$#2$}][#1]{$#3$}\else\makebox[\widthof{#2}][#1]{#3}\fi}
\definecolor{darkblue}{rgb}{0.0,0.0,0.7}
\definecolor{darkred}{rgb}{0.7,0.0,0}
\definecolor{lightblue}{rgb}{0.6, 0.7, 1}
\definecolor{lightred}{rgb}{1, 0.8, 0.8}
\definecolor{lightgray}{gray}{0.75}

\usepackage[pdftex,
        colorlinks=true,
        linkcolor=darkblue,
        citecolor=darkblue,
        filecolor=black,
        pagecolor=black,
        urlcolor=darkblue,
        bookmarks=true,
        bookmarksopen=true,
        bookmarksopenlevel=3,
        plainpages=false,
        pdfpagelabels=true,
        verbose=true]{hyperref}

\usepackage[all]{hypcap}

\begin{document}

\title{Engineering Ising-XY spin models in a triangular lattice\\via tunable artificial gauge fields}

\author{J. Struck}

\author{M. Weinberg}

\author{C. \"Olschl\"ager}

\author{P. Windpassinger}

\author{J. Simonet}
\affiliation{\footnotesize Institut f\"ur Laserphysik, Universit\"at Hamburg, Luruper Chaussee 149, D-22761 Hamburg, Germany}

\author{K. Sengstock}
\affiliation{\footnotesize Institut f\"ur Laserphysik, Universit\"at Hamburg, Luruper Chaussee 149, D-22761 Hamburg, Germany}
\affiliation{\footnotesize Zentrum  f\"ur Optische Quantentechnologien, Universit\"at Hamburg, Luruper Chaussee 149, D-22761 Hamburg, Germany}

\author{\\R. H\"oppner}
\affiliation{\footnotesize Institut f\"ur Laserphysik, Universit\"at Hamburg, Luruper Chaussee 149, D-22761 Hamburg, Germany}
\affiliation{\footnotesize Zentrum  f\"ur Optische Quantentechnologien, Universit\"at Hamburg, Luruper Chaussee 149, D-22761 Hamburg, Germany}

\author{P. Hauke}
\affiliation{\footnotesize Institut f\"ur Quantenoptik und Quanteninformation, \"Osterreichische Akademie der Wissenschaften, A-6020 Innsbruck, Austria}

\author{A. Eckardt}
\affiliation{\footnotesize Max-Planck-Institut f\"ur Physik komplexer Systeme, N\"othnitzer Str.\,38, D-01187 Dresden, Germany}

\author{M. Lewenstein}
\affiliation{\footnotesize Institut de Ci\`encies Fot\`oniques, Av. Carl Friedrich Gauss 3, E-08860 Castelldefels, Barcelona, Spain}
\affiliation{\footnotesize ICREA-Instituci\`o Catalana de Recerca i Estudis Avan\c{c}ats, Lluis Companys 23, E-08010 Barcelona, Spain}

\author{L. Mathey}
\affiliation{\footnotesize Institut f\"ur Laserphysik, Universit\"at Hamburg, Luruper Chaussee 149, D-22761 Hamburg, Germany}
\affiliation{\footnotesize Zentrum  f\"ur Optische Quantentechnologien, Universit\"at Hamburg, Luruper Chaussee 149, D-22761 Hamburg, Germany}

\maketitle
\textbf{Emulation of gauge fields for ultracold atoms provides access to a class of exotic states arising in strong magnetic fields. Here we report on the experimental realisation of tunable staggered gauge fields in a periodically driven triangular lattice. For maximal staggered magnetic fluxes, the doubly degenerate superfluid ground state breaks both a discrete $\mathbb{Z}_2$ (Ising) symmetry and a continuous $U(1)$ symmetry.
\newline
By measuring an Ising order parameter, we observe a thermally driven phase transition from an ordered antiferromagnetic to an unordered paramagnetic state and textbook-like magnetisation curves. Both the experimental and theoretical analysis of the coherence properties of the ultracold gas demonstrate the strong influence of the $\mathbb{Z}_2$ symmetry onto the condensed phase.}
\newline

Phase transitions in systems with combined continuous and discrete symmetries are fundamentally different from their purely continuous and discrete counterparts. The interplay between different types of excitations in the various degrees of freedom can lead to a complex behaviour and coupling of the associated order parameters \cite{Villain1977b, Yosefin1985,Choi1985,SachdevBook, DiepBook}. A paradigm example is the fully frustrated XY model on a triangular lattice. It combines vector spin-type symmetries with discrete chiral degrees of freedom, which result in the famous spin--chirality coupling at low temperatures \cite{Hasenbusch2005}. However, experimental studies in solid-state systems are challenging in view of implementing and isolating an XY model Hamiltonian \cite{Ling1996,Martinoli2000,Affolter2002}.

Ultracold bosonic quantum gases in optical lattices, on the other hand, constitute a highly versatile system with an extraordinary degree of control \cite{Bloch2008, Lewenstein2012}. In particular, the recent experimental realisations of artificial gauge potentials for bulk \cite{Cornell2004,Dalibard2004,Spielman2009,Spielman2011} and optical lattice systems \cite{Bloch2011,Spielman2012,Struck2011,Struck2012} allow for the investigation of new physical regimes, not realisable in condensed matter systems.
\begin{figure}
\centering
\hypertarget{fig:Figure1}{}
\includegraphics{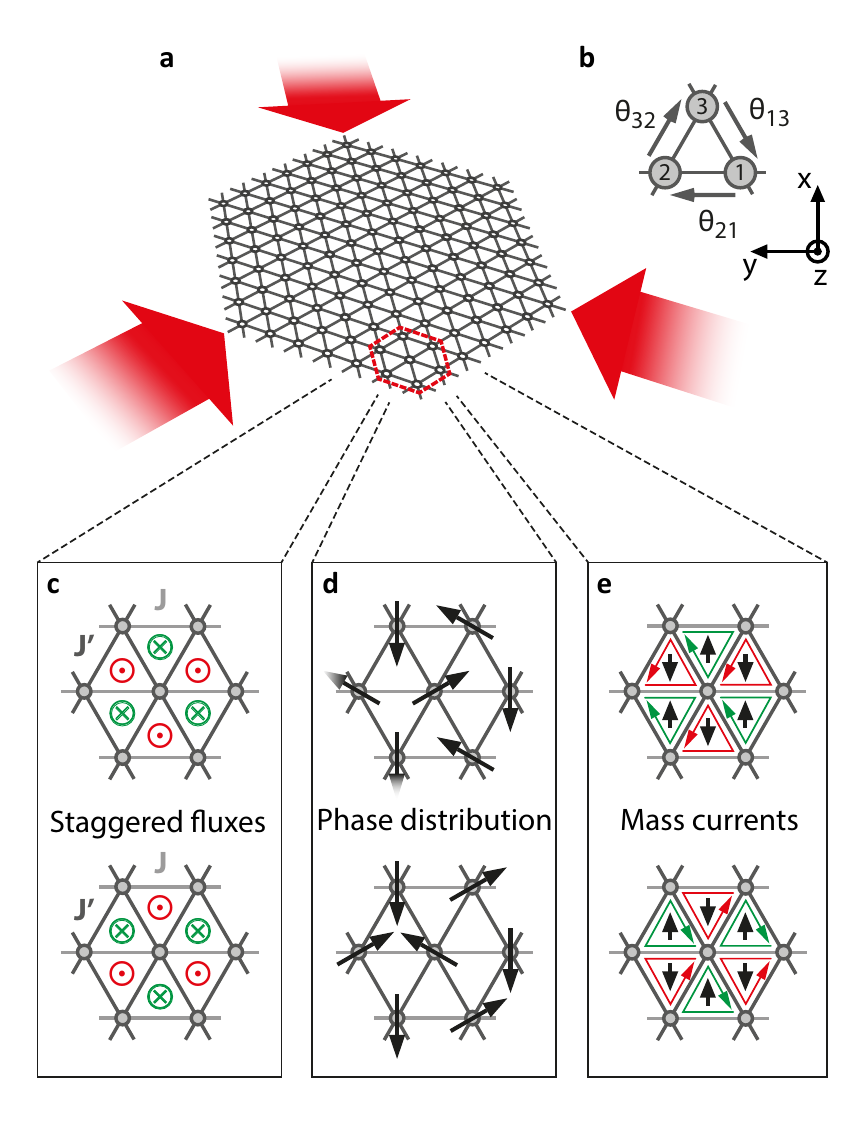}
\caption{\textbf{Illustration of the triangular optical lattice, the artificial gauge fluxes, the phase distribution, and the mass currents.} \textbf{a}, The triangular optical lattice is created by the interference of three running-wave laser beams. In the experiment, roughly 2000 triangular plaquettes are occupied. \textbf{b}, Orientation of the arguments of the tunneling matrix element around a plaquette. \textbf{c}, A strong artificial, staggered gauge field is applied to the lattice system. Crosses (dots) correspond to inwards (outwards) pointing gauge fluxes. \textbf{d}, An accumulated flux of $\pm\pi$ around neighbouring plaquettes results in two, energetically degenerate, phase configurations. These phase configurations lead to opposite chiralities in the mass currents around the plaquettes. \textbf{e}, The current on each plaquette defines the orientation of an Ising-type spin.}
\label{Figure1}
\end{figure}

Here, we demonstrate the realisation of a system with combined $U(1)$ and $\mathbb{Z}_2$ symmetries using ultracold atoms submitted to artificial gauge fields. Our experimental setup consists of an ultracold gas of $^{87}$Rb atoms held in a two-dimensional triangular lattice \cite{Becker2010} (see Fig.\,\hyperlink{fig:Figure1}{\ref{Figure1}a}). At each lattice site $j$ with particle number $N_j$, the weakly interacting superfluid gas can be described by the local order parameter $\langle a_j\rangle =\sqrt{N_j}e^{\mathrm{i}\varphi_j}$. As a central aspect, the local phases $\varphi_j$ are mapped onto classical XY spins $\mathbf{s}_j=(\cos{\varphi_j},\sin{\varphi_j})$, where the tunneling matrix elements between neighbouring lattice sites correspond to the spin-spin coupling parameters. Such classical spins possess a continuous degree of freedom. In presence of a long-range order, analogous to the onset of Bose-Einstein condensation (BEC), the order parameter assumes an arbitrary, but fixed phase, thus breaking the continuous $U(1)$ symmetry \cite{PitaevskiiBook}.

Beyond that, we experimentally engineer strong staggered gauge fields, which generate an additional discrete $\mathbb{Z}_2$ symmetry in our system. The resulting magnetic flux induces cyclotron-like mass currents around each plaquette. The two possible chiralities of these currents circulating around a single plaquette correspond to a discrete Ising-like order parameter. Furthermore, the tunability of the artificial gauge fields enables us to bias the $\mathbb{Z}_2$ order parameter, in analogy to a longitudinal external magnetic field in the Ising-spin model.

The result is a flexible model system which allows us to study the temperature-dependent behaviour and interplay of the discrete and continuous order parameters.

In the work presented here, the complex tunneling matrix elements, necessary to generate staggered fluxes, are created by accelerating the lattice potential along a closed orbit. A suitable periodic forcing \cite{Struck2011,Arimondo2012} results in the following effective Bose-Hubbard Hamiltonian:
\begin{equation}
H_{\text{eff}}=-\sum_{\left\langle i,j \right\rangle}{|J_{ij}| e^{\mathrm{i}\theta_{ij}} a_i^\dag a_j}+ \frac{U}{2} \sum_{j}{n_j(n_j-1)}
\end{equation}
where the spatial degrees of freedom perpendicular to the lattice have been omitted for clarity (see Supplementary Material).
Here, $a_j^\dag$ ($a_j$) is the creation (annihilation) operator of a boson at lattice site $j$, $n_j\,{=}\,a_j^{\dag}a_j$ is the respective number operator, and $U$ is an on-site repulsion. In the kinetic term, the summation over the nearest neighbours is directional as $\theta_{ji}\,{=}\,{-}\theta_{ij}$.
The hopping parameters along the directions $2\rightarrow3$ and $3\rightarrow1$ (see Fig.\,\hyperlink{fig:Figure1}{\ref{Figure1}b}) are equal and denoted as $|J'| e^{\mathrm{i}\theta'}$ in the following. Experimentally, $|J_{21}| e^{\mathrm{i}\theta_{21}}\,{\equiv}\,|J| e^{\mathrm{i}\theta}$ and $|J'| e^{\mathrm{i}\theta'}$  can be tuned independently of each other. The total phase accumulated on a closed path around one triangular plaquette reflects the gauge flux through the cell, defined as  $\Phi\,{\equiv}\,\Phi_{\bigtriangleup}\,{=}\,(\theta\,{+}\,2\theta')\,\mathrm{mod}\,2\pi\,{=}\,{-}\,\Phi_{\bigtriangledown}$. The global acceleration of the lattice potential realised here induces fluxes with opposite sign for upwards and downwards pointing plaquettes, as depicted in Fig.\,\hyperlink{fig:Figure1}{\ref{Figure1}c}.

The triangular lattice is fully frustrated for staggered fluxes of maximum magnitude $\pi$. For this extreme case the flux structure is not unique (since $-\pi\,\mathrm{mod}\,2\pi\,{=}\,\pi\,\mathrm{mod}\,2\pi$) and the two flux patterns sketched in Fig.\,\hyperlink{fig:Figure1}{\ref{Figure1}c} are equivalent. This equivalence leads to two energetically degenerate vector spin configurations, as depicted in Fig.\,\hyperlink{fig:Figure1}{\ref{Figure1}d}. The staggered currents induced by the gauge fluxes display the same degeneracy (see Fig.\,\hyperlink{fig:Figure1}{\ref{Figure1}e}). Note that both the argument $\theta_{ij}$ of the tunneling parameter and the relative orientation $(\varphi_j-\varphi_i)$ of the XY spins influence the mass current $\langle j_{ij}\rangle$ along one lattice bond:
\begin{eqnarray}
\langle j_{ij}\rangle&=&-\frac{2|J_{ij}|}{\hbar} \mathrm{Im}(e^{\mathrm{i}\theta_{ij}} \langle a_i^\dag a_j\rangle)\\
&=&-\frac{2|J_{ij}|}{\hbar}\sqrt{N_iN_j} \sin(\theta_{ij}+\varphi_j-\varphi_i).
\end{eqnarray}
The strong interplay between the chirality of the cyclotron-like mass currents (Ising parameter) and the XY spin long-range order induces the coupling between the broken $\mathbb{Z}_2$ and $U(1)$ symmetries in our system.

\begin{figure*}
\centering
\hypertarget{fig:Figure2}{}
\includegraphics{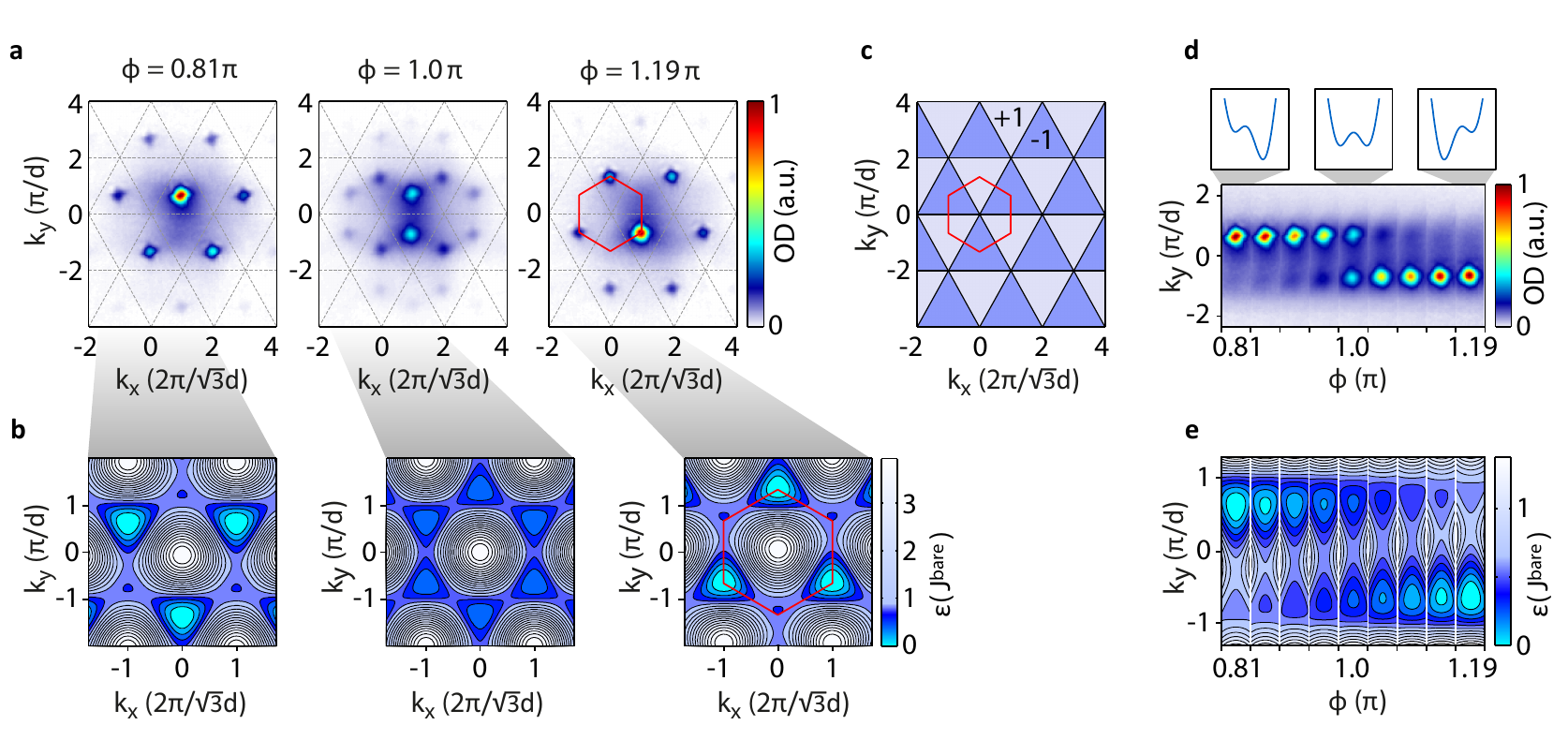}
\caption{\textbf{Effect of staggered gauge fluxes in momentum space.} \textbf{a}, Experimentally observed occupations of the momentum states within the lowest Bloch band in TOF images, averaged over about 200 single-shot realisations, and  \textbf{b} calculated dispersion relations for the given values of the gauge flux $\Phi$. The red hexagon indicates the first Brillouin zone. \textbf{c}, The value of the real-space Ising order parameter corresponds to the occupation of a triangular mask in quasimomentum space. Atoms in the +1 ($-1$) regions correspond to positive (negative) chirality. \textbf{d}, A zoom into the central region along $k_y$ of the TOF images shows the relative occupation of the two Ising modes as a function of the gauge flux. The observation is in good agreement with the position and the relative importance of the minima in the band structure as shown in \textbf{e}.}
\label{Figure2}
\end{figure*}

The presence of staggered gauge fluxes has a direct signature in the momentum space. The single-particle dispersion relation of the lattice is indeed strongly deformed:
\begin{eqnarray}
\begin{split}
\varepsilon (\mathbf{k}) \,{=}\, &-2|\phantomas{J'}{J}|\cos\big( \mathbf{k}\cdot \mathbf{a}_1 -\theta\phantom{'}\big) \\
                                 &-2|J'|               \cos\big( \mathbf{k}\cdot \mathbf{a}_2 -\theta' \big)\\
                                 &-2|J'|               \cos\big( \mathbf{k}\cdot \mathbf{a}_3 -\theta' \big)
\end{split}
\end{eqnarray}
where the $\textbf{a}_i$ are the lattice directions (see Methods). For  $\Phi=\pi$, it exhibits two degenerate minima with opposite $k_y$ values within the first Brillouin zone, while for fluxes of $\Phi=\pi \pm \beta$ this degeneracy is lifted (see Fig.\,\ref{Figure2}).
For ultracold bosonic gases, the changes in the momentum space occupation can be easily observed with standard time-of-flight (TOF) imaging techniques, where the in-situ quasimomentum distribution is converted into position information. Figure\,\hyperlink{fig:Figure2}{\ref{Figure2}a} shows TOF images summed over many experimental realisations for three amplitudes of the staggered gauge fluxes. For $\Phi=\pi$, both momentum modes are equally populated on average. For a strong bias flux of $\beta= 0.2\,\pi$, we externally drive the system into one of the minima in the first Brillouin zone. The corresponding dispersion relations, plotted in Fig.\,\hyperlink{fig:Figure2}{\ref{Figure2}b}, illustrate the deformations induced by the different values of the gauge fluxes. Figure\,\hyperlink{fig:Figure2}{\ref{Figure2}d-e} demonstrates the experimental control over the degeneracy between the two minima in the first Brillouin zone. In analogy to the effect of a longitudinal magnetic field in the Ising model, the ability of tuning the flux strength $\Phi$ thus enables us to bias the system towards one of the two minima.

The measured quasimomentum distribution contains in fact more information, reflecting both symmetries of the system.
A long-range order of the XY spins, which breaks the $U(1)$ symmetry, implies that the momentum distribution is singular.

The two possible chiralities of the mass currents correspond to quasimomenta in complementary parts of the Brillouin zone (see Supplementary Material). Measuring the differential occupation in the two momentum classes, depicted as upwards and downwards pointing triangles in Fig.\,\hyperlink{fig:Figure2}{\ref{Figure2}c}, gives access to the mean chirality of the system. This analogue to the Ising magnetisation is analysed in the following.

\begin{figure*}
\centering
\hypertarget{fig:Figure3}{}
\includegraphics{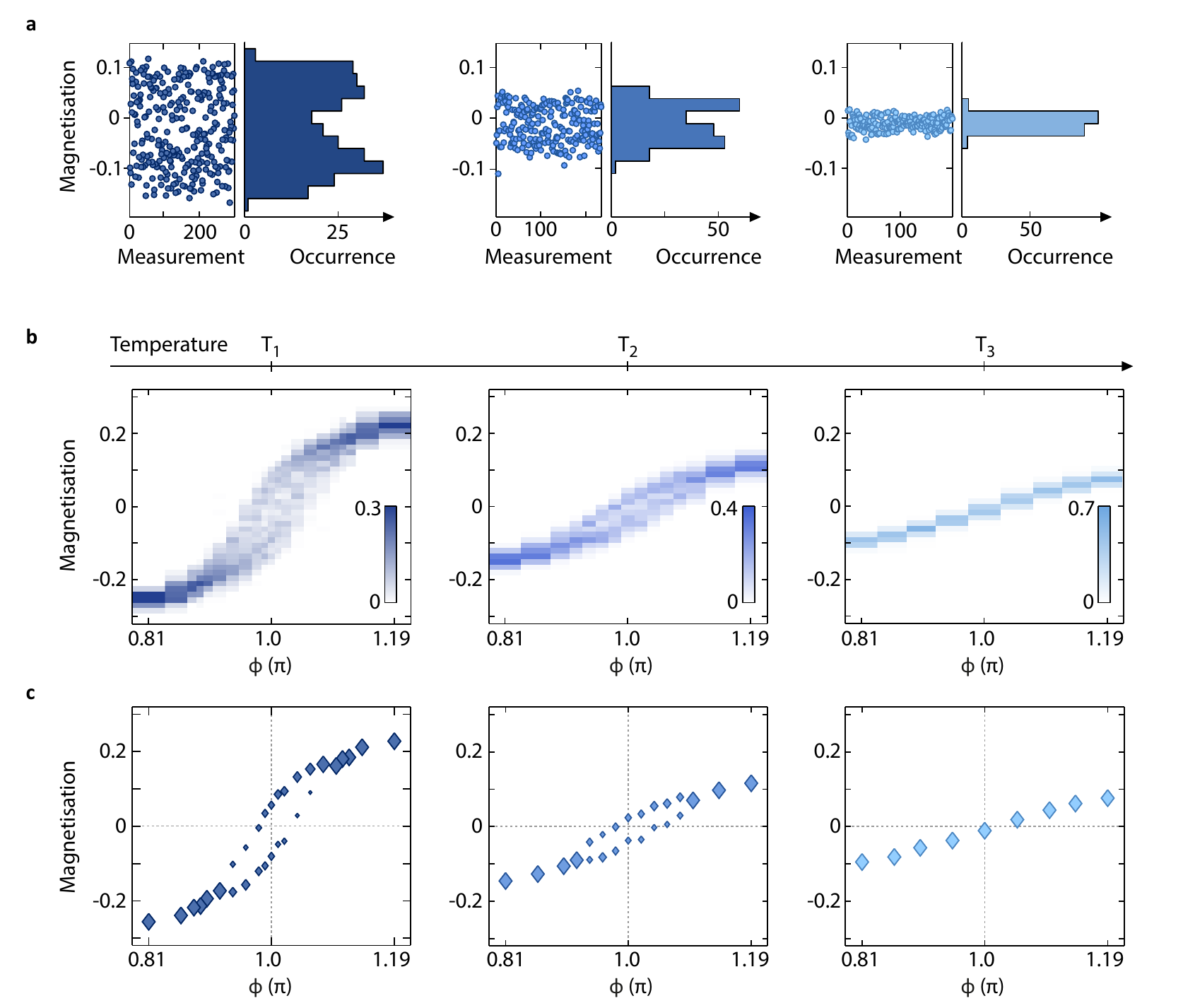}
\caption{\textbf{Measurement of the statistical distribution of the chiral magnetisation.} \textbf{a}, The statistical distribution of the magnetisation $(\bigtriangleup-\bigtriangledown)/(\bigtriangleup+\bigtriangledown)$ obtained from consecutive single experimental realisations at flux strength $\Phi\,{=}\,\pi$ (left) and the corresponding histograms (right) are shown for three different temperatures. \textbf{b}, Histograms revealing the statistical distribution of the sample magnetisation for different temperatures and gauge fluxes. For each of these histograms about 200 individual measurements have been recorded. The colour code corresponds to the normalised amplitudes of the histograms. \textbf{c}, Maxima of Gaussian probability distributions which are fitted to the raw data. For bimodal distributions, the point size represents their relative weighting.}
\label{Figure3}
\end{figure*}

As a central result, a thermally induced phase transition between an antiferromagnetic and a paramagnetic phase can be observed. Figure\,\hyperlink{fig:Figure3}{\ref{Figure3}a} shows a statistical analysis of consecutive, individual experimental realisations for $\Phi=\pi$ for three different temperatures. For individual measurements, the Ising-type magnetisation fluctuates. At the lowest temperature achieved, its statistical distribution clearly shows the spontaneous breaking of the $\mathbb{Z}_2$ symmetry into two individual modes. When the temperature is increased, the spontaneous magnetisation decreases and finally vanishes when the system crosses the phase boundary to an unordered paramagnetic state. The simultaneous observation of both Ising states in a single experimental realisation is very likely due to spatial phase separation of different chiralities similar to the formation of magnetic Weiss domains.

The bias flux $\beta$ impacts onto the occupation of the two Ising states. In Fig.\,\hyperlink{fig:Figure3}{\ref{Figure3}b-c}, the measurement of the magnetisation as a function of the gauge flux in the three temperature regimes is presented. For each value of the gauge flux, the statistical distribution of the magnetisation is represented by normalised histograms in row b. Row c shows the maxima of a uni- or bimodal probability distribution fitted to the raw data (see Supplementary Material).

For a large bias flux $\beta$ the system is completely magnetised in one of the two Ising states as expected for an Ising spin system subjected to a longitudinal magnetic field. Below the critical temperature and in the vicinity of flux $\Phi=\pi$, we can identify two branches of favored magnetisations which correspond to the occupation of the two Ising states.
This behaviour cannot be explained for a system in thermal equilibrium. Indeed, already the presence of a small external magnetic field suppresses the condensation in the state with higher energy. However, this state corresponds to a local minimum and the system can become metastable. The finite occupation probability of the excited Ising state stems from non-adiabatic dynamics. The amplitude of the artificial gauge field is progressively increased to its final value. During this preparation ramp the dispersion relation becomes flat and thus the energy barrier between the two states is increased from almost zero to the final value (see Supplementary Material for more details). Therefore, a finite probability exists for the system to be trapped in the local minimum.

The experimentally observed metastability arises from the repulsive interactions between the atoms, which prevents fragmentation of the state. This is supported by theoretical calculations including these interactions.
Namely, the free energy of the system can be evaluated up to the first-order correction in the interaction strength \cite{Huang1957} (see Supplementary Material). At low temperatures, the effective free energy develops two minima. Condensation in one of the two minima is equivalent to the spontaneous breaking of the $\mathbb{Z}_2$ symmetry. The energy scale protecting the metastable minimum is the mean-field energy, which is large compared to the temperature. This conclusion is also confirmed by a Bogoliubov-de Gennes theory, including the second-order correction with respect to interactions (i.e. the quantum fluctuations) (see Methods). For this reason, we can observe the metastable states.

At higher temperatures the entropic contribution to the free energy merges the two minima into one. Therefore, no metastable state is expected above the $\mathbb{Z}_2$ critical point. The measured occupation of metastable states with a magnetisation opposite to the bias field is therefore a non-equilibrium signature of the phase transition.

\begin{figure}
\centering
\includegraphics{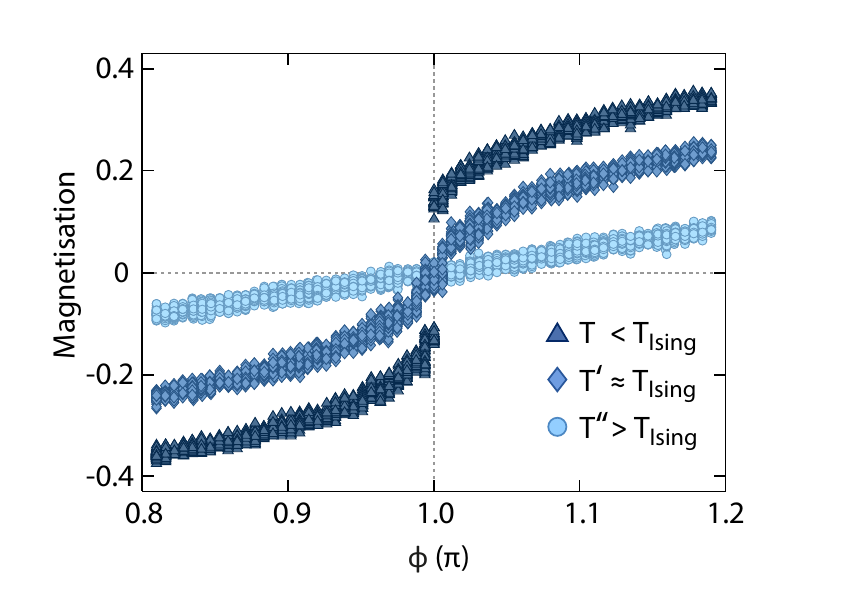}
\hypertarget{fig:Figure4}{}
\caption{\textbf{Magnetisation curves obtained via a classical Monte Carlo simulation.} At the lowest temperature ($T$), a non-zero spontaneous magnetisation is well reproduced, which disappears as the temperature is increased ($T' < T''$). $T_\mathrm{Ising}$ denotes the critical temperature for the $\mathbb{Z}_2$ symmetry breaking. The apparent scattering reflects the fluctuations of a series of MC simulations.}
\label{Figure4}
\end{figure}
The equilibrium state of the studied three-dimensional system can be investigated via a classical Monte Carlo (MC) approach. Here, all relevant experimental parameters, including the overall confinement, have been taken into account in the simulation (see Supplementary Material). Figure\,\ref{Figure4} shows the magnetisation curves for three different temperatures, generated by extracting the chirality from the momentum distribution. For flux $\Phi=\pi$, the thermally driven phase transition from an ordered state showing spontaneous magnetisation to an unordered state is reproduced, and overall similar to the experimental data. No finite occupation of the metastable minimum can be observed, since the MC simulation generates the ground state of the system.

\begin{figure}
\centering
\hypertarget{fig:Figure5}{}
\includegraphics{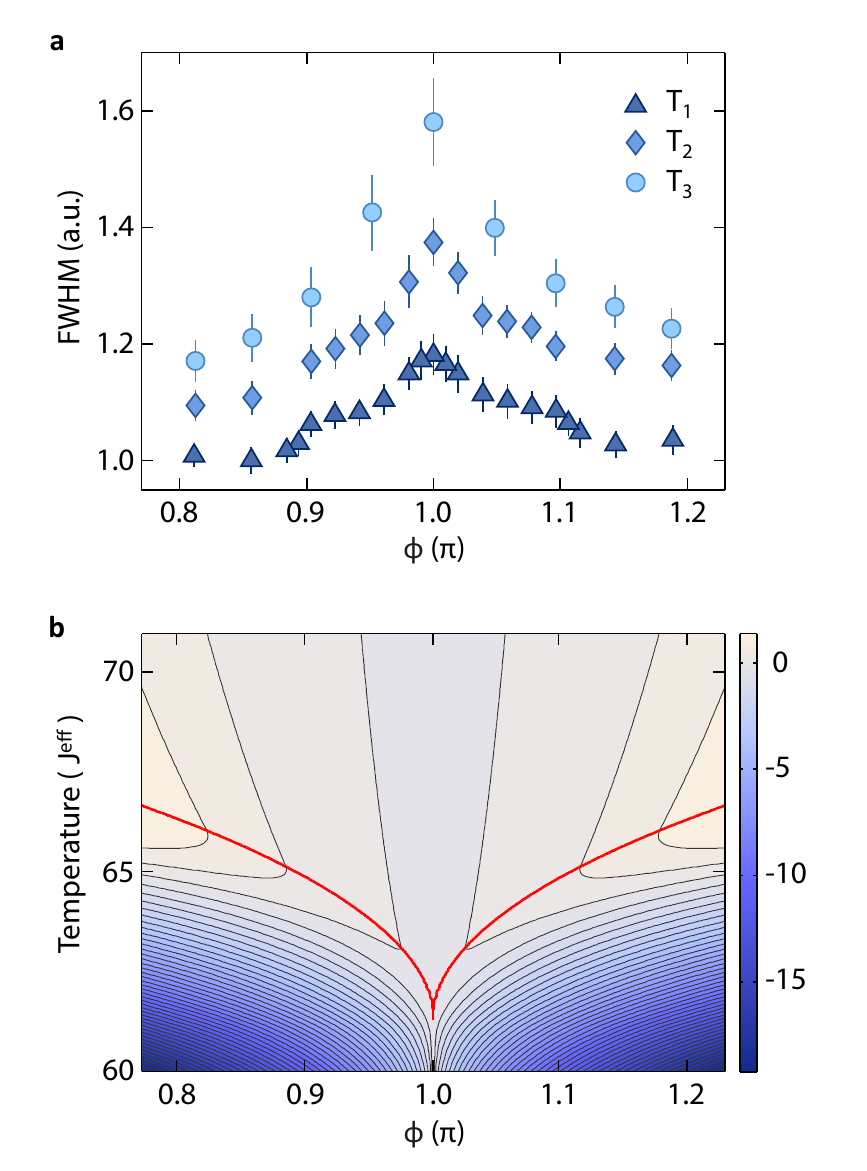}
\caption{\textbf{Experimental and theoretical evaluations, related to the $U(1)$ order parameter.} \textbf{a}, Measured FWHM of the momentum peaks for the three different temperatures $T_1<T_2<T_3$. The width of the momentum peaks is a measure for the loss of long-range coherence of the sample. \textbf{b}, Free energy in units of $J^\mathrm{eff}/\mu m^3$ as a function of flux and temperature. The condensation transition (red line) shows a cusp at $\pi$-flux. }
\label{Figure5}
\end{figure}
While the statistical distribution of the magnetisation quantifies the $\mathbb{Z}_2$ symmetry breaking, the sharpness of the momentum peaks is a measure for the long-range phase coherence connected to the $U(1)$ symmetry. The peak full-width-half-maximum (FWHM), extracted from the experimental data presented in Fig.\,\hyperlink{fig:Figure3}{\ref{Figure3}b-c}, is shown in Figure \hyperlink{fig:Figure5}{\ref{Figure5}a}. As expected, for each flux value the peak width increases with the temperature, monitoring the decreasing long-range order. More remarkably the $U(1)$ order parameter depends strongly on the gauge flux strength. For a deeper understanding of this behaviour, the critical temperature for BEC ($T_c$)  has been theoretically evaluated in the weak-coupling approximation of the free energy  and is plotted in Fig.\,\hyperlink{fig:Figure5}{\ref{Figure5}b} (see Supplementary Material). For measurements realised at a fixed temperature, the coherence length of the gas should decrease in the vicinity of $\Phi=\pi$, where $T_c$ displays a pronounced cusp. The measured full-width-half-maximum is limited by the finite time-of-flight, but the short coherence lengths expected in the vicinity of $\Phi=\pi$ are nicely reproduced. The observed increase of the measured FWHM symmetric to $\Phi=\pi$ is in good agreement with the theory.
Similar conclusions follow from the exact thermodynamic analysis of the non-interacting gas (see Supplementary Material).

In conclusion, we have realised a model system with Ising-type $\mathbb{Z}_2$ and global $U(1)$ phase symmetry by applying strong gauge fields to bosonic atoms in a triangular optical lattice. For classical two-dimensional XY systems with coupled spin and chirality degrees of freedom, theory predicts that the system first breaks the $\mathbb{Z}_2$ chiral symmetry and then the $U(1)$ symmetry as the temperature is reduced \cite{Korshunov2002}. However, the exact nature of these phase transitions, which are strongly linked by combined excitations, has long been debated \cite{Lee1998,Capriotti1998}. Only recently, precise Monte Carlo simulations could resolve the two transitions and identify their universality classes \cite{Okumura2011,Obuchi2012}. The analysis of the coherence properties of the 3D ultracold gas demonstrates the strong influence of the $\mathbb{Z}_2$ symmetry breaking onto the BEC phase, revealed as a drastic reduction of the coherence length. In future, it will be interesting to investigate the coupling between these phase transitions and its influence on their critical behaviour, which is however experimentally challenging. In addition, the occupation of metastable states with a magnetisation opposite to the bias field is a non-equilibrium signature of the Ising-like phase transition. This constitutes a fundamental, defining property of such phase transitions, which is observed here in the field of ultracold atoms.

This work paves the way to further studies of artificial magnetic properties of ultracold quantum gases in optical lattices. Combinations of the two-dimensional control of the complex tunneling parameters reported here with superlattices \cite{Hauke2012} in different lattice geometries (triangular, hexagonal, or kagome) promise to give deeper insights into a variety of magnetic systems \cite{Lewenstein2012,DiepBook}. \newline

We acknowledge support from the Deutsche Forschungsgemeinschaft (GRK1355, SFB925) and the Landesexzellenzinitiative Hamburg (supported by the Joachim Herz Stiftung), ERC AdG QUAGATUA, AAII-Hubbard, Spanish MICINN (FIS2008-00784), Catalunya-Caixa, EU Projects AQUTE and NAMEQUAM, the Spanish foundation Universidad.es, the Austrian Science Fund (SFB F40 FOQUS), the DARPA OLE program and the John von Neumann Institute for Computing (NIC) for providing us with computing time on the supercomputers of the Juelich Supercomputing Centre (JSC).

\section{Methods}

\scalefont{.82}

\textbf{The triangular optical lattice.} The two dimensional, triangular optical lattice as depicted in Fig.\,\hyperlink{fig:Figure1}{1d} is created by three running laser beams with actively stabilised phases that intersect in the $xy$-plane at an angle of $120^{\circ}$. The beams are derived from a Ti:sapphire laser at wavelength $\lambda_L\,{=}\,830\,\mathrm{nm}$, creating a 2D lattice potential $V{(\mathbf{r})}\,{=}\,{-}\,V_0\sum_i\cos(\mathbf{b}_i\mathbf{r})$ with a lattice spacing of $d\,{=}\,2\lambda_L/3\,{=}\,533\,\mathrm{nm}$. The reciprocal lattice directions are $\mathbf{b}_1\,{=}\,b/2(1,\sqrt{3},0)$, $\mathbf{b}_2\,{=}\,b(1,0,0)$ and $\mathbf{b}_3\,{=}\,b/2(-1,\sqrt{3},0)$, where $b\,{=}\,2\pi\sqrt{3}/\lambda_L$, corresponding to the real-space lattice directions $\mathbf{a}_1\,{=}\,d(0,\,1,\,0)$, $\mathbf{a}_2\,{=}\,d/2(\sqrt{3},\,-1,\,0)$ and $\mathbf{a}_3\,{=}\,{-}d/2(\sqrt{3},\,1,\,0)$.
\\\hspace{5mm}

\textbf{Experimental preparation.} We create Bose-Einstein condensates of $(1.5-2.5)\times10^5$ $^{87}\mathrm{Rb}$ atoms in a crossed optical dipole trap. Within $100\,\mathrm{ms}$, we subsequently ramp up the optical lattice to a final lattice depth of $(4.6\pm0.1)\,\mathrm{E}_{\mathrm{rec}}$ ($\mathrm{E}_{\mathrm{rec}}=h\times3.33\,\mathrm{kHz}$) which leads to a bare single-particle tunneling parameter of $J^{\mathrm{bare}}\,{=}\,4\times10^{-3}\,\mathrm{E}_{\mathrm{rec}}$. As the system is only weakly confined in $z$-direction with respect to the lattice potential -- the overall external harmonic confinement is $\boldsymbol\omega_{\mathrm{Tot}}\,{=}\,2\pi\times(31,53,40)\,\mathrm{Hz}$ -- the atoms form an array of approximately $2100$ to $2600$ elongated tubes with a mean occupancy in the range of $70$ to $95$ atoms ($175$ to $235$ in the center). The temperature of the system is increased by holding the atoms longer in the lattice before applying the artificial gauge fluxes.

Note that the system under study is three-dimensional. Therefore we observe a BEC transition instead of a Kosterlitz Thouless transition as expected for a pure two-dimensional system.
\\\hspace{5mm}

\textbf{Lattice shaking.} Staggered fluxes in the triangular lattice are realised by a global periodic motion of the optical lattice around a closed orbit $\mathbf{x}(t)\,{=}\,-A_x\cos(\tilde{\omega}t)\mathbf{\hat{e}}_x\,{-}\,A_y\left[\sin(\tilde{\omega} t)\,{+}\,\delta\sin(2\tilde{\omega}t)/4\right]\mathbf{\hat{e}}_y$, with $\tilde{\omega}\approx2\pi\times2.8\,\mathrm{kHz}$. The amplitude of the staggered flux can be accessed by the control parameter $\delta$. In the reference frame of the moving lattice, this results in a force
\begin{equation}
\mathbf{F}(t) = - m \ddot{\mathbf{x}}(t) = -F_x\cos(\tilde{\omega}t)\mathbf{\hat{e}}_x-F_y\left[\sin(\tilde{\omega}t)+\delta\sin( 2\tilde{\omega}t)\right]\mathbf{\hat{e}}_y
\end{equation}
acting on the atoms. Experimentally, the trajectory is realised by modulating two of the three lattice laser beams with $\nu_{2/3}\,{=}\,{\pm}\,\nu_x\sin(\tilde{\omega} t)\,{+}\,\nu_y\left[\cos(\tilde{\omega}t)\,{+}\,\delta\cos(2\tilde{\omega}t)/2\right]$, where $\nu_x\,{=}\,\tilde{\omega}A_x/(\sqrt{3}d)$,  $\nu_y\,{=}\,\tilde{\omega}A_y/d$ and $A_{x,y}\,{=}\,F_{x,y}/(m\tilde{\omega}^2)$. Hereby, the shaking amplitudes $\nu_x$ and $\nu_y$ are linearly increased to their final values in $50\,\mathrm{ms}$ after the condensate is loaded into the initially resting lattice. Time-averaging the projection of the force onto the bonds of the elementary plaquette now leads to a renormalisation of the tunneling matrix elements $J^{\mathrm{bare}}\rightarrow J^{\mathrm{eff}}$ in the $xy$-plane \cite{Eckardt2005}. The absolute values of the effective tunneling matrix elements are on the order of $|J^{\text{eff}}|\,{\approx}\,0.4\,J^{\text{bare}}$.
\\\hspace{5mm}

\textbf{Symmetries of the system.} With $U/J^{\mathrm{eff}}=1.2-1.4$ and large filling factors, the system remains in the weakly interacting regime. Here, the local wavefunction has well defined phases $\varphi_j$ on each lattice site, which correspond to the $XY$ vector spins $\mathbf{s}_j\,{=}\,\left(\cos\varphi_j, \sin\varphi_j\right)$. In the case where the arguments of the effective hopping are equal to $\pi$ (i.e. $\delta\,{=}\,0$), the total kinetic energy of the atomic ensemble along the lattice directions can be written as
\begin{align}
E(\lbrace\varphi_i\rbrace) &= \sum_{\langle i,j\rangle}|J^{\mathrm{eff}}_{ij}|\cos(\varphi_j-\varphi_i)\notag\\
&=\sum_{\langle i,j\rangle}|J^{\mathrm{eff}}_{ij}|\mathbf{s}_i\cdot\mathbf{s}_j.
\label{eqn:SpinEnergy}
\end{align}
Equation (\ref{eqn:SpinEnergy}) is invariant under both a discrete $\mathbb{Z}_2$ transformation and a global $U(1)$ rotation:
\begin{equation}
\mathbf{s}_j \rightarrow \mathbf{s}^\prime_j =
\begin{cases}
 \big(\cos(\varphi_j), - \sin(\varphi_j)\big) & \mathbb{Z}_2  \\
\big(\cos(\varphi_j+ \nu), \sin(\varphi_j + \nu)\big) & U(1)
\end{cases}
\end{equation}
On the contrary, the chirality changes its sign under the discrete transformation. The summation of this quantity over the lattice plaquettes corresponds to the magnetisation of the system.
\\\hspace{5mm}

\textbf{Detection and data analysis.} All the information about the momentum distribution of the superfluid are retrieved from absorption images taken after $32\,\mathrm{ms}$ time-of-flight. The chirality is defined by the spins at the corners of one elementary plaquette as $\chi\,{=}\,\mathrm{sgn}\left[\mathbf{s}_{2}\times\mathbf{s}_{1}+\mathbf{s}_{3}\times\mathbf{s}_{2}+\mathbf{s}_{1}\times\mathbf{s}_{3}\right]$, where $\mathbf{s}_{i}\times\mathbf{s}_{j}\,{\equiv}\, s_{i,x}s_{j,y}\,{-}\,s_{i,y}s_{j,x}$. It can be converted to a mask in quasimomentum space (see Fig.\,\hyperlink{fig:Figure2}{2c}). By weighting each absorption image with this mask we obtain the total magnetisation of the system as shown in Fig.\,\hyperlink{fig:Figure3}{3b}.
\\\hspace{5mm}

\textbf{Bogoliubov theory for the metastable condensate.} In a translational invariant system with a Bose-condensate in one of the two local minima of the free dispersion relation $\varepsilon(\mathbf{k})$, one can add quantum and thermal fluctuations within Bogoliubov theory. One obtains the quasiparticle dispersion relation
\begin{align}
\omega(\mathbf{q})
=
\frac{ \tilde{\varepsilon}(\mathbf{q})-\tilde{\varepsilon}(-\mathbf{q})}{2}
+ \sqrt{  \left[ g_{_{\text{1D}}} \rho_{_{\text{1D}}}
+\frac{ \tilde{\varepsilon}(\mathbf{q})+\tilde{\varepsilon}(-\mathbf{q})}{2} \right]^2 - g^2_{_{\text{1D}}} \rho^2_{_{\text{1D}}}  }
\label{eqn:bogo}
\end{align}
for momenta $\mathbf{q}=\mathbf{k}-\mathbf{k}_0$ relative to the condensate momentum $\mathbf{k}_0$, where $\tilde{\varepsilon}(\mathbf{q}) = \varepsilon(\mathbf{k}_0+\mathbf{q})-\varepsilon(\mathbf{k}_0)$ and with $g_{_{\text{1D}}}$ and  $\rho_{_{\text{1D}}}$ denoting the interaction parameter and the density in the tubes respectively. A thermodynamic instability is indicated when $\omega(\mathbf{q})$ assumes negative values for some $\mathbf{q}$. For vanishing interaction $g_{_{\text{1D}}}\rho_{_{\text{1D}}} = 0$, one has $\omega(\mathbf{q})= \tilde{\varepsilon}(\mathbf{q})$ and the system is thermodynamically unstable as soon as the condensate is not prepared in the global minimum of the dispersion relation. However, finite interactions $g_{_{\text{1D}}}\rho_{_{\text{1D}}} > 0$ can stabilise a condensate in the upper local minimum of the dispersion with $\omega(\mathbf{q})> 0$. The very same mechanism leads to spontaneous symmetry breaking for $\Phi\,{=}\,\pi$ by disfavoring a fractionalised condensation. The existence of a metastable state is thus directly linked to spontaneous symmetry breaking.

\newpage

\scalefont{1.22}

\part{Supplementary material}
\hspace{1mm}

\setcounter{figure}{0}
\setcounter{table}{0}
\setcounter{equation}{0}
\renewcommand{\thefigure}{S\arabic{figure}}
\renewcommand{\thetable}{S\arabic{table}}
\renewcommand{\theequation}{S\arabic{equation}}

\section{Lattice shaking}

\begin{figure}[b]
\hypertarget{fig:s1}{}
\includegraphics{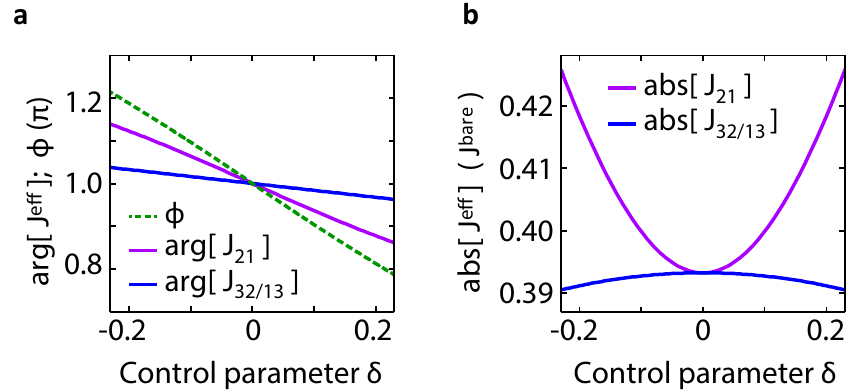}
\caption{\textbf{Tunability of the complex tunneling parameters.} \textbf{a}, The Peierls phases along the bonds and the resulting flux strength through an elementary plaquette are plotted as a function of the control parameter $\delta$. \textbf{b}, Magnitudes of the three effective tunneling matrix elements in units of the bare tunneling amplitude.}
\label{FluxPhaseTunnelingDelta}
\end{figure}

As stated in [\hyperlink{link:SuppBib}{S1}], staggered fluxes in triangular lattices can be realised by a global periodic motion of the optical lattice around a closed orbit. The trajectory used for the experiments presented in this article is given by\hypertarget{eq:s1}{}
\begin{equation}
\begin{split}
\mathbf{x}(t) \,{=}\,   &- A_{x} \cos(\tilde{\omega} t) \mathbf{e}_x \\
                        &- A_{y} \left[\vphantom{A^2} \sin(\tilde{\omega} t) + \delta\sin(2\tilde{\omega} t)/4\right] \mathbf{e}_y,
\label{eq:LatticeTrajectory}
\end{split}
\end{equation}
where $\tilde{\omega}\,{=}\, 2 \pi / T \,{=}\,2 \pi\,{\times}\, 2.791\,\mathrm{kHz}$ with $\mathbf{x}(t) \,{=}\, \mathbf{x}(t+T)$.
The important control parameter for the staggered flux strength is $\delta$. For $\delta\,{=}\,0$ all tunneling matrix elements are real valued and only flux strengths which are zero or $\pi$ can be achieved.
The inertial force acting on the atoms in the reference frame of the moving lattice is
\begin{equation}
\mathbf{F}(t) \,{=}\, - F_x \cos(\tilde{\omega} t) \mathbf{e}_x - F_y \left[\vphantom{A^2} \sin(\tilde{\omega} t) + \delta \sin(2 \tilde{\omega} t)\right] \mathbf{e}_y,
\label{eq:LatticeForcing}
\end{equation}
where the connection to the trajectory (eqn.\,\hyperlink{eq:s1}{S1}) is given by $A_{x} \,{=}\, F_{x}/(m \tilde{\omega}^2)$ and $A_{y} \,{=}\, F_{y}/(m \tilde{\omega}^2)$. Experimentally the forcing of the atoms in the lattice is realised by frequency modulating two of the three lattice laser beams with
\begin{equation}
\begin{split}
\Delta \nu_{2/3} \,{=}\, \pm &\nu_{x} \sin(\tilde{\omega} t) \\
    + &\nu_{y} \left[ \vphantom{A^2} \cos(\tilde{\omega} t) + \delta\cos(2 \tilde{\omega} t)/2 \right],
\end{split}
\end{equation}

where $\nu_{x} \,{=}\, F_{x}/(\sqrt{3} d m \tilde{\omega})$ and $\nu_{y} \,{=}\, F_{y}/(d m \tilde{\omega})$.
The renormalised tunneling matrix elements due to the time averaging over one cycle are\hypertarget{eq:s4}{}
\begin{equation}
J_{ij}^{\mathrm{eff}} \,{=}\,  \frac{J^{\mathrm{bare}}}{T} \int_{0}^{T} \mathrm{d}t \exp\left(\vphantom{A^2}\mathrm{i} W_{ij}(t) / \hbar\right),
\label{eq:TunnelMatrixRenormalisation}
\end{equation}
with
\begin{equation}
W_{ij}(t) \,{=}\,- \int_{-\infty}^{t} \mathrm{d} \tau \mathbf{F}(\tau) ~ \mathbf{a}_{j}.
\end{equation}
The vectors $\mathbf{a}_{j}$ describe a closed path around one elementary plaquette of the lattice. Taking advantage of the symmetries in our system, the tunneling matrix elements are written as $J\,{=}\,J_{21}$ and $J'\,{=}\,J_{32}\,{=}\,J_{13}$ in the following.

Fig.\,\hyperlink{fig:s1}{S1a} depicts the numerical solutions for the Peierls phases and the resulting staggered flux according to the equation (\hyperlink{eq:s4}{S4}). In Fig.\,\hyperlink{fig:s1}{S1b} the magnitude of the different effective tunneling matrix elements are shown. Note that the difference between the magnitude of the tunneling matrix elements is on the order of a few percents. Therefore, it only has a weak influence on the dispersion, as will be detailed in the next section.

\begin{figure}
\hypertarget{fig:s2}{}
\centering
\includegraphics{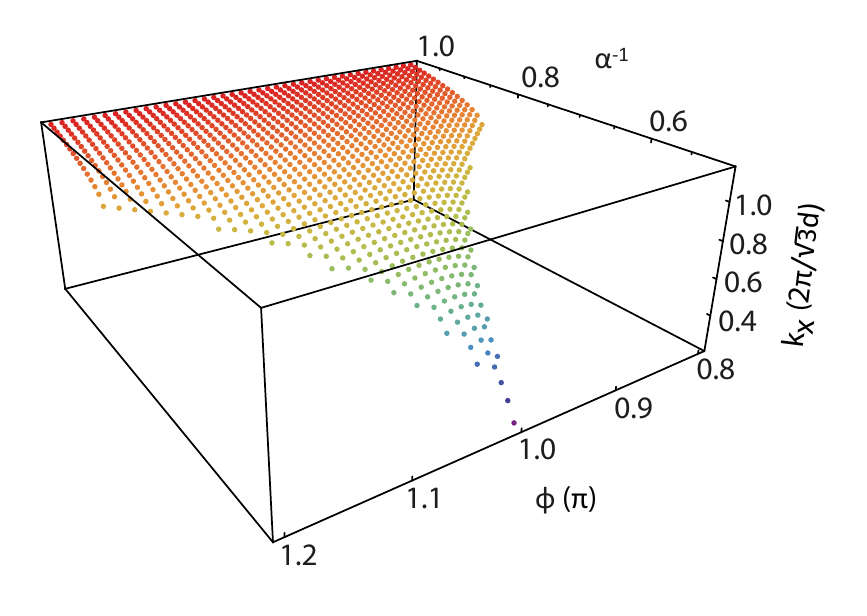}
\caption{\textbf{Minima in the first Brillouin zone in presence of gauge fluxes.} k-space separation between the ground- and excited state quasi momenta in dependence of the anisotropy parameter $\alpha$ and the staggered flux strength $\Phi$.}
\label{kDifference}
\end{figure}

\section{The dispersion relation}
The lowest band dispersion relation for the triangular lattice in presence of complex tunneling matrix elements is described by
\begin{equation}
\begin{split}
\varepsilon (\mathbf{k}) \,{=}\, &-2|\phantomas{J'}{J}|\cos\big( d k_{y} - \theta_{21}\big) \\
                                 &-2|J'|               \cos\big( d \left[\smash{\sqrt{3}}k_{x}  - k_{y} \right]/2 - \theta_{32} \big)\\
                                 &-2|J'|               \cos\big( d \left[\smash{\sqrt{3}}k_{x}  + k_{y} \right]/2 + \theta_{13} \big).
\end{split}
\end{equation}
The gauge invariant quantity of the system is the staggered flux strength $\Phi$. On the contrary, the hopping arguments $\theta_{ij}$ depend on the chosen gauge. A change of gauge corresponds to a translation of the band structure. The specific choice of the gauge $\theta_{21}\,{=}\,\pi\,{+}\,\beta$, $\theta_{32}\,{=}\,\pi$ and $\theta_{13}\,{=}\,\pi$ yields the simplified expression for the dispersion relation
\begin{equation}
\begin{split}
\varepsilon (\mathbf{k}) \,{=}\, +&2|\phantomas{J'}{J}| \cos\left(\vphantom{A^2}d k_{y} - \beta \right)\\
                                 +&4|J'|                \cos\left(\vphantom{A^2}d k_{x} \smash{\sqrt{3}}/2\right) \cos\left(\vphantom{A^2}d k_{y} /2\right).
\label{eq:GaugeDispersionRelation}
\end{split}
\end{equation}
The $x$-component of the quasi-momentum for the local minimum is $q_{\text{min},x} \,{=}\, 2 \pi/(d \sqrt{3})$. For $\Phi\,{=}\, \pi$ the corresponding $y$-component can be written as:
\begin{equation}
q_{\text{min},y}\,{=}\,\begin{cases}
0  & \mathrm{for} ~\alpha>2\\
\pm \frac{2}{d} \arccos\left( \frac{\alpha}{2} \right)  & \mathrm{for} ~\alpha<2\\
\end{cases}
\end{equation}
where $\alpha\,{=}\, J/J'$ is the anisotropy parameter of the lattice. The numerical results for the quasi-momentum separation between ground-state minimum and metastable minimum are shown in Fig.\,\hyperlink{fig:s2}{S2}. Since in our case the anisotropy parameter remains close to unity, the quasi-momentum separation is only weakly depending on the flux.

\begin{figure}
\hypertarget{fig:s3}{}
\includegraphics{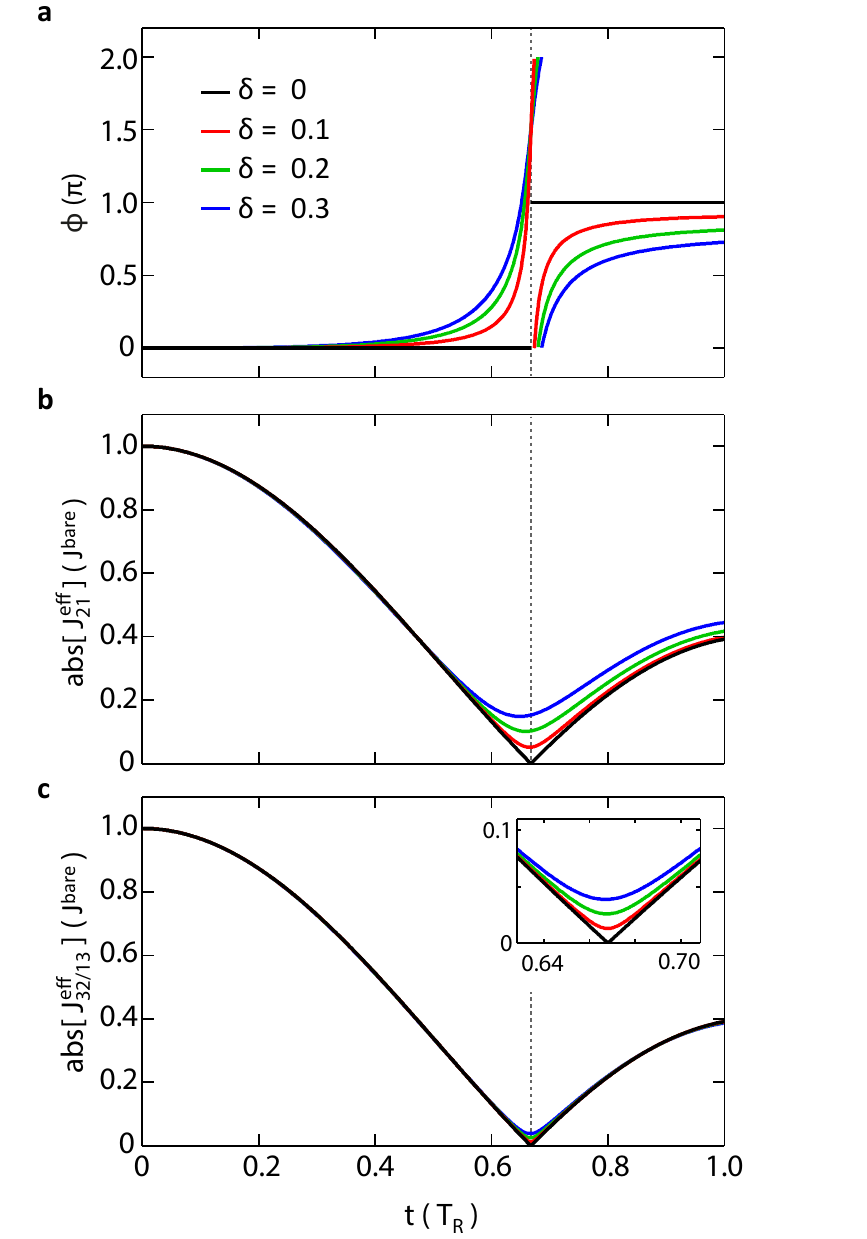}
\caption{\textbf{Switching of the gauge fluxes.} \textbf{a}, Amplitudes of the gauge flux strength and \textbf{b}, \textbf{c} of the tunneling matrix elements along the bonds $1 \rightarrow 2$ and $2\rightarrow3$ $(3\rightarrow1)$ respectively are plotted for different values of the control parameter $\delta$ as a function of the time during the linear ramping of $\nu_{x}$ and $\nu_{y}$. The time is expressed in units of the ramp time $T_R$. The inset in \textbf{c} is a zoom into the region of small absolute tunneling matrix elements $J_{32/13}$.}
\label{Ramping}
\end{figure}

\section{Switching the gauge fluxes}
After the lattice potential has been ramped to its final depth of $4.6\,E_{\text{rec}}$, the tunneling matrix elements are all real and positive valued. In order to induce non-zero gauge fluxes, the frequency modulation of the laser beams is slowly turned on by increasing the frequency amplitudes $\nu_{x}$ and $\nu_{y}$ linearly over a time $T_R\,=\,50\,\mathrm{ms}$. Depending on the value of the control parameter $\delta$, staggered gauge fluxes with different final amplitude can be realised.

Since the ramping time scale is slow compared to the orbital motion of frequency $\tilde{\omega}\,{=}\,2 \pi\,{\times}\, 2.791\,\mathrm{kHz}$, the system has well defined tunneling matrix elements during the switching procedure. The time resolved evolution of the phases and amplitudes of the effective hopping elements during the ramping of $\nu_{x}$ and $\nu_{y}$ are shown in Fig.\,\hyperlink{fig:s3}{S3}. It is important to note that the fluxes are rapidly switched to their final amplitude. This corresponds to a quench into the final state and explains the non-adiabatic behaviour described in the main text. As observed experimentally, slow ramps reduce the excitations in the system but since the absolute tunneling values become small during the ramp (see Fig.\,\hyperlink{fig:s3}{S3b} and \hyperlink{fig:s3}{c}), the process cannot be fully adiabatic in experimentally accessible time scales.

The initial temperature of the system has been varied by holding the atomic sample in the lattice prior to introducing the staggered gauge fluxes by shaking. The chosen durations were 0\,ms, 80\,ms and 160\,ms respectively for the three investigated regimes.

\section{Tube parameters}

In order to understand the physical properties of the complete system in all three dimensions, it is important to derive some basic parameters for the array of elongated tubes in $z$-direction that are formed by the presence of the optical lattice in the $xy$-plane. The basic ansatz for the wavefunction of the system is
\begin{equation}
\psi(\mathbf{r}) \,{=}\, \sum_i c_i w_i(x,y) \zeta_i(z),
\end{equation}
where $w_i(x,y)$ is the single particle Wannier function of the 2D lattice, $\zeta_i(z)$ is an interaction broadened function along the tubes and $N_i\,{=}\,|c_i|^2$ is the number of atoms in the tube of lattice site $i$. Neglecting the kinetic energy of the system, this leads to a Gross-Pitaevskii equation for the functions $\zeta_i(z)$. The squared modulus of the function is given by
\begin{equation}
|\zeta_i(z)|^2 \,{=}\, \frac{\mu - m \left(\omega_{x}^2 R_{i,x}^2 + \omega_{y}^2 R_{i,y}^2 + \omega_{z}^2 z^2 \right)/2}{\tilde{g}N_{i}},
\end{equation}
where the $\omega_{x,y,z}$ denote the overall external harmonic confinement and $R_{i,x}, R_{i,y}$ are the $x$,$y$ components of the lattice vector of site $i$. $\tilde{g}\,{=}\,g \int \mathrm{d}x \mathrm{d}y |w_i(x,y)|^4$ is the renormalised interaction parameter with the bare three dimensional interaction parameter $g\,{=}\, 4 \pi \hbar^2 a_{s} /m$. For $^{87}\mathrm{Rb}$: $a_{F\,{=}\,2}\,{=}\, + (100.4 \pm 0.1) a_{0}$ [\hyperlink{link:SuppBib}{S2}]. A numerical calculation of the 2D-Wannier functions yields the result $\tilde{g}/g \,{=}\, 17.2\,{\times}\, \mathrm{\mu m}^{-2}$ for  $4.6\,\mathrm{E_{rec}}$. The length of the tubes $(2 z_{\mathrm{TF}})$ is determined by the Thomas-Fermi boundaries in $z$-direction
\begin{equation}
z_{\mathrm{TF}} \,{=}\, \sqrt{\frac{2 \mu - m \left( \omega_{x}^2 R_{i,x}^2 + \omega_{y}^2 R_{i,y}^2 \right)/2}{m \omega_{z}^2}}
\end{equation}
and the number of particles in a single tube is $\smash{N_{i}\,{=}\, 2 m \omega_{z}^2 z_{\mathrm{TF}}^{3}/(3\tilde{g})}$. By making the continuum approximation $\sum_{i}N_i\,{\rightarrow}\,A_{\mathrm{UC}}^{-1}\int\mathrm{d}x\mathrm{d}yN(x,y)$ with $R_{i,x},R_{i,y}\,{\rightarrow}\,x,y$, the chemical potential can be calculated analytically:
\begin{equation}
\mu \,{=}\, \left( \frac{15\tilde{g} A_{\mathrm{UC}} }{16 \pi \sqrt{2}} \, \omega_{x} \, \omega_{y} \,  \omega_{z} \, m^{3/2} \, N_{\mathrm{Tot}} \right)^{2/5},
\end{equation}
where $A_{\mathrm{UC}} \,{=}\, \sqrt{3/4} ~ d^2$ is the area of the unit cell and $N_{\mathrm{Tot}}$ the total particle number of the system. The energy scale associated with each tube $i$ can be calculated as $U_i\,{=}\, \tilde{g}\int \mathrm{d}z |\zeta_i(z)|^4 \,{=}\, 3\tilde{g}/(5z_{\mathrm{TF}})$. Relevant system parameters for total particle numbers of $1.5\times10^5$ and $2.5\times10^5$ are depicted in Tab.\,\ref{tab:TubeParameters}.

\section{Time-of-flight measurements}
After rapidly switching off all trapping potentials and letting the atoms fall in free space for $32\,\mathrm{ms}$ we take an absorption image of the cloud with a magnification of approximately $3$. With this standard time-of-flight method, the quasi-momentum distribution of the atoms in the lattice can be revealed. In Fig.\,\hyperlink{fig:s4}{S4}, samples of averaged time-of-flight images are shown in dependence of flux strength and temperature in the lattice. As stated before, the physics of the shaken system is described by a time-averaged effective Hamiltonian with renormalised tunneling matrix elements. The quasi-momentum distribution describing the effective model is static, while the only effect of the fast periodic acceleration is an overall oscillating envelope on top of this quasi-momentum distribution. The density distribution in the far-field regime after time-of-flight is given by
\begin{equation}
n(\mathbf{k}) \,{=}\, \left|\widetilde{w}_0\left(\mathbf{k}-\frac{m\dot{\mathbf{r}}}{\hbar}\right)\right|^2\sum_{i,j}{e^{i\mathbf{k}(\mathbf{R_i}-\mathbf{R_j})}\left<{a}^\dagger_{i}{a}^{\vphantom{\dagger}}_{j}\right>},
\end{equation}
where $\widetilde{w}_0$ is the Fourier-transformed Wannier function driven by the shaking.  The expectation value in the sum describes the coherence properties of the sample. In order to keep the same Wannier envelope position, the switch-off time is chosen to occur at the same time within one period for all measurements. The center is positioned in between the two degenerate minima (for $\Phi=\pi$) of the dispersion relation at $\mathbf{k_{C}}\,{=}\,(+2\pi/\sqrt{3}d,\,0)$. However, a small displacement remains towards $k_y\,{<}\,0$. For the given envelope size the displacement leads to a slightly favored weighting of the negative magnetisation in the performed measurements. This is the reason for the negative offset of the data presented in Fig.\,\ref{Figure3}.

\begin{table}
\centering
\renewcommand{\arraystretch}{1.5}
{\setlength{\fboxsep}{0pt}%
\colorbox{lightgray!25}{%
\begin{tabular}{lll}
\specialrule{0.5mm}{0pt}{0pt}
$\vphantom{\bigg[}N_{\mathrm{Tot}}$                                          & $1.5 \,{\times}\, 10^5$  & $2.5 \,{\times}\, 10^5$ \\ \hline
$\vphantom{\Big[}N_{\mathrm{Sites}}$                                        & 2157              &  2629 \\
$\vphantom{\Big[}N_{\mathrm{Tot}}/N_{\mathrm{Sites}} $                      & 70                & 95 \\
$\vphantom{\Big[}N_{\mathrm{max}}$                                          & 174               & 237 \\
$\vphantom{\Big[}U/J \,{=}\, \sum_{i} (U_{i} \, N_{i})/(N_{\mathrm{Tot}} \, 0.4 \, J^{\mathrm{bare}})\hspace{6mm}$ & 1.4 & 1.2 \\
$\vphantom{\Big[}l_{\mathrm{Tube}} = \sum_{i} (2 \, z_{i,\mathrm{TF}} \, N_{i})/N_{\mathrm{Tot}}$ & $22.7\,\mu\mathrm{m}$ & $25.2\,\mu\mathrm{m}$ \\
$\vphantom{\Big[}\rho_{\mathrm{1D}} \,{=}\, \sum_{i} \left[N_{i}^2/(2 \, z_{i,\mathrm{TF}})\right] / N_{\mathrm{Tot}} $ & $4.6\,\mu\mathrm{m}^{-1}$ &  $5.6\,\mu\mathrm{m}^{-1}$ \vspace{1mm}\\
\specialrule{0.5mm}{0pt}{0pt}
\end{tabular}}}
\flushleft
\caption{\textbf{Calculated system parameters for different total particle numbers.} Parameters are: the number of occupied sites $N_{\mathrm{Sites}}$, mean tube occupancy $N_{\mathrm{Tot}}/N_{\mathrm{Sites}}$, maximum tube occupancy $N_{\mathrm{max}}$, occupation weighted ratio $U/J$, occupation weighted tube length $l_{\mathrm{Tube}}$ and occupation weighted 1D-Density $\rho_{\mathrm{1D}}$. }
\label{tab:TubeParameters}
\end{table}

\begin{figure*}[t]
\hypertarget{fig:s4}{}
\includegraphics{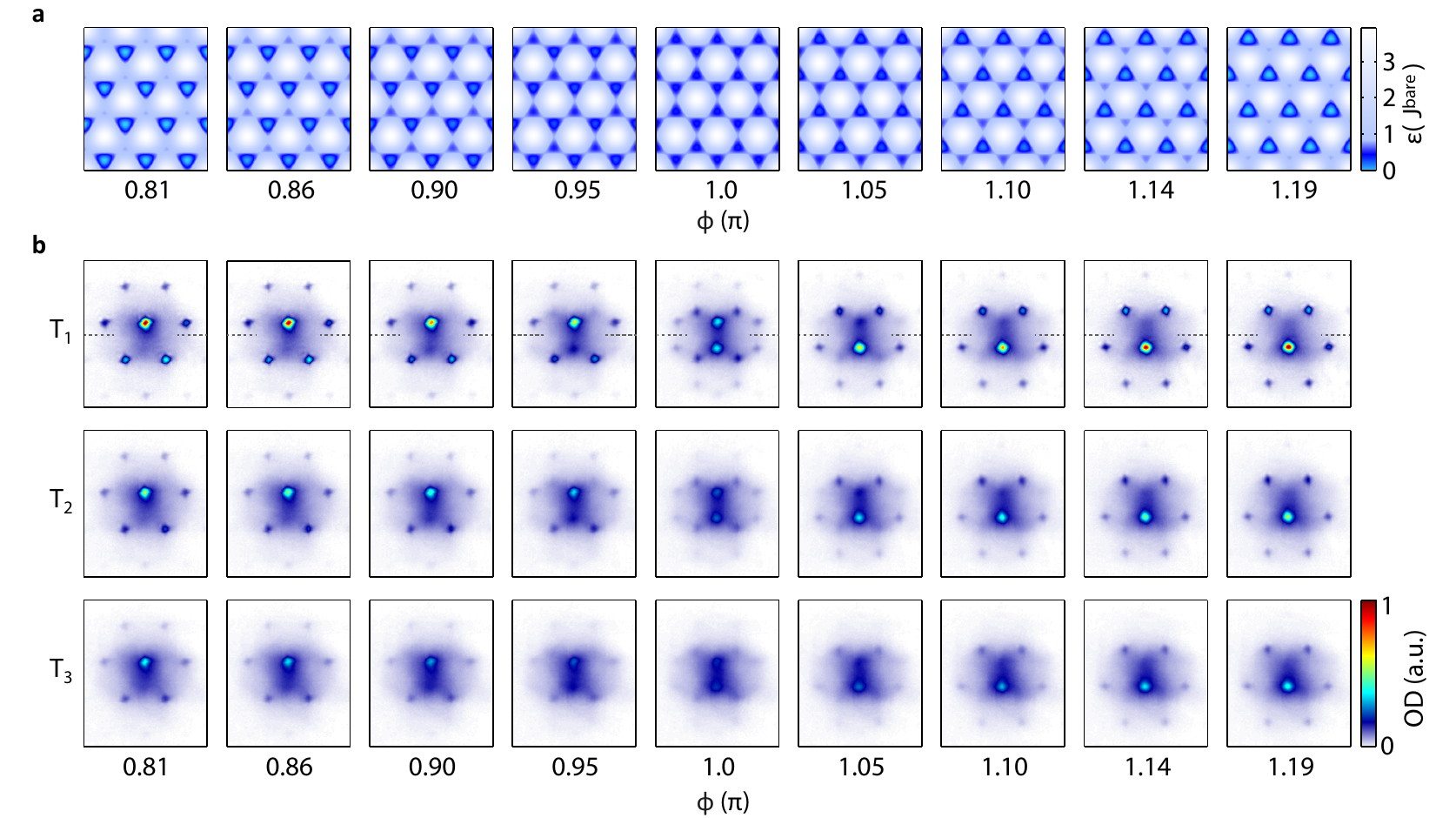}
\caption{\textbf{Quasi-momentum distributions.} \textbf{a}, Dispersion relation $\varepsilon(\mathbf{k})$ for selected values of the flux strength $\Phi$. \textbf{b}, The corresponding averaged time-of-flight images for the three different initial temperatures $T_1$, $T_2$ and $T_3$ show the characteristic population of the minima in the dispersion relation. The dashed lines in the first row of images are a guide to the eye.}
\label{Tofimages}
\end{figure*}

\section{Gauge-independent chirality masks}
In order to determine the magnetisation of the system from the time-of-flight images, we introduce the chirality of the system
\begin{equation}
\chi \,{=}\, \mathrm{sgn} \big[\vphantom{A^2} \mathbf{s}_{2} \,{\times}\, \mathbf{s}_{1} + \mathbf{s}_{3} \,{\times}\, \mathbf{s}_{2} + \mathbf{s}_{1} \,{\times}\, \mathbf{s}_{3} \big], \label{eq:chiralitymask}
\end{equation}
where the scalar cross product of our two dimensional vector spins is defined as
\begin{equation}
\begin{split}
\mathbf{s}_{i} \times \mathbf{s}_{j} &\equiv  \epsilon^{ij}\mathbf{s}_i\mathbf{s}_j\\
                                     &=  s_{i,x} s_{j,y} - s_{i,y} s_{j,x}.
\end{split}
\end{equation}
The spins $\mathbf{s}_{1},\mathbf{s}_{2}, \mathbf{s}_{3}$ are arranged clockwise around a triangular plaquette. For staggered mass currents this results in the same value of $\chi$ for the two types of triangular plaquettes (upwards- and downwards pointing triangles in $x$-direction). The chirality can be converted into the reciprocal space, resulting in a mask for the TOF images
\begin{equation}
\label{eq:chiralitymask reciprocal space}
\tilde{\chi}(\mathbf{k}) \,{=}\, \mathrm{sgn}\left[\sum_{i=1}^3\sin(\mathbf{k}\cdot\mathbf{a}_{i})\right].
\end{equation}
\begin{figure}
\hypertarget{fig:s5}{}
\centering
\includegraphics{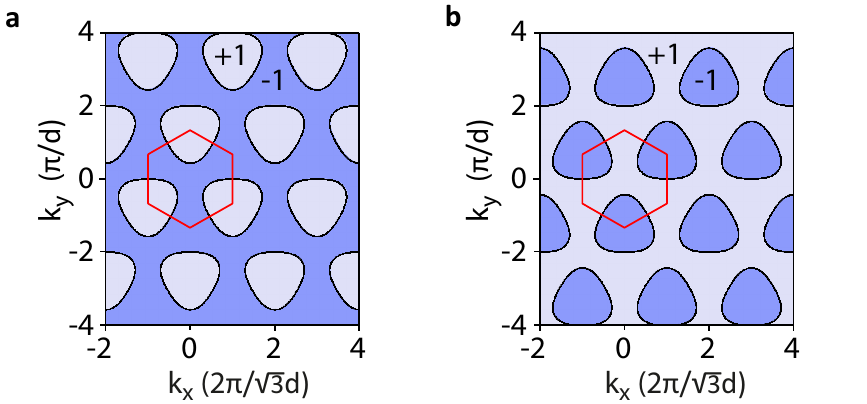}
\caption{\textbf{Gauge dependency of the chirality masks.} The chirality masks $\tilde{\chi}_G$ obtained from the sign of equation (\ref{eq:chiralitymasktrue}) are plotted for different gauges: \textbf{a} $\theta\,{=}\,0.9\pi$, $\theta'\,{=}\,0.95\pi$ and \textbf{b} $\theta\,{=}\,1.1\pi$, $\theta'\,{=}\,1.05\pi$. The first Brillouin zone is indicated by a red hexagon.}
\label{Chiralitytrue}
\end{figure}
Weighting the TOF images of $n(\textbf{k})$  with this mask gives access to the mean chirality or, in the Ising picture, the staggered magnetisation of the system $M\,{=}\,\int \tilde{\chi}(\textbf{k})n(\textbf{k})\mathrm{d}^2\text{k}$. In principle this quantity is not exact since it is gauge dependent. An observable which characterises the $\mathbb{Z}_2$ order parameter in a gauge independent way is given by the total staggered flux $\mathcal{J_\text{Tot}}$, that is
\begin{equation}
\begin{split}
\langle\mathcal{J_\text{Tot}}\rangle\,&{=}\,\sum_i^{N_{\text{sites}}}\sum_{j=1}^3\langle j_{i\textbf{a}_j}\rangle\\
&{=}\,\frac{2|J|}{\hbar}\sum_{\textbf{k}}\langle n\left(\textbf{k}\right)\rangle\mathcal{X}\left(\textbf{k},\theta,\theta'\right).
\end{split}
\end{equation}
Here, $n\left(\textbf{k}\right)$ is the density in momentum space and $\langle j_{i\textbf{a}_j}\rangle$ denotes the expectation value of the mass current from lattice site $j$ to the nearest neighbour in the direction $\textbf{a}_j$. With the different $\theta$ and $\theta'$, the specific gauge that was chosen is described by the weighting function
\begin{equation}
\label{eq:chiralitymasktrue}
\begin{split}
\mathcal{X}\left(\textbf{k},\theta,\theta'\right)\,{=}\,\phantom{+}&\sin\left(\textbf{k}\cdot\textbf{a}_1-\theta\phantom{'}\right)\\
+&\sin\left(\textbf{k}\cdot\textbf{a}_2-\theta'\right)\\
+&\sin\left(\textbf{k}\cdot\textbf{a}_3-\theta'\right).
\end{split}
\end{equation}
It can be used to define a set of gauge-independent chirality masks $\tilde{\chi}_G\left(\textbf{k},\theta,\theta'\right)\,{=}\,\text{sgn}\left[\mathcal{X}\left(\textbf{k},\theta,\theta'\right)\right]$. The resulting masks for the specific gauge of $\theta\,{=}\,0.9\pi$, $\theta'\,{=}\,0.95\pi$ and $\theta\,{=}\,1.1\pi$, $\theta'\,{=}\,1.05\pi$, that correspond to flux strengths of $\Phi\,{=}\, 0.8\pi$ and $\Phi\,{=}\,1.2\pi$, respectively, are shown in Fig.\,\hyperlink{fig:s5}{S5}. For simplicity, we use the mask defined by equation (\ref{eq:chiralitymask reciprocal space}), where $\theta\,{=}\,\theta'\,{=}\,\pi$, and $\tilde{\chi}_G\,{=}\,\tilde{\chi}$, since the difference of the magnetisation data generated with gauge independent masks turns out to be negligibly small.

\begin{figure}
\hypertarget{fig:s6}{}
\centering
\includegraphics{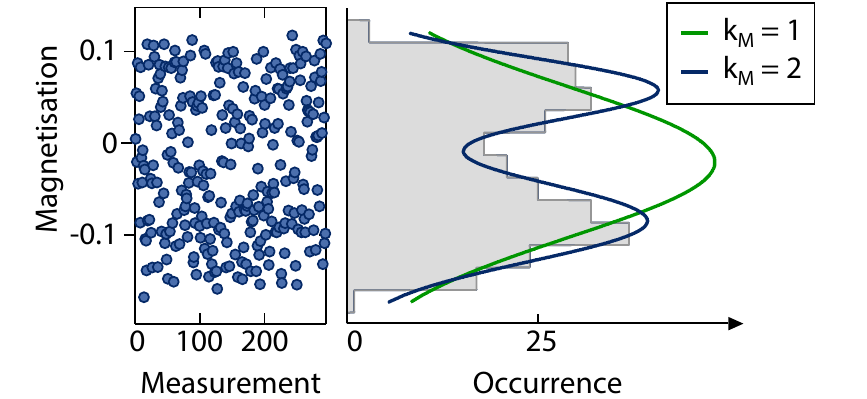}
\flushleft
\caption{\textbf{Analysis of the statistical distribution of the magnetisation.} Illustration of the bimodal fluctuation of the magnetisation due to the breaking of the $\mathbb{Z}_2$ symmetry for the case of low temperatures and flux $\Phi\,{=}\,\pi$ as seen in Fig.\,\ref{Figure3}\textcolor{darkblue}{a}. The solid lines on the right represent the fitted uni- $(k_M\,{=}\,1)$ and bimodal $(k_M\,{=}\,2)$ Gaussian probability distributions. }
\label{ClusteringExample}
\end{figure}

\section{Statistical data analysis}

\begin{figure*}
\centering
\hypertarget{fig:s7}{}
\includegraphics{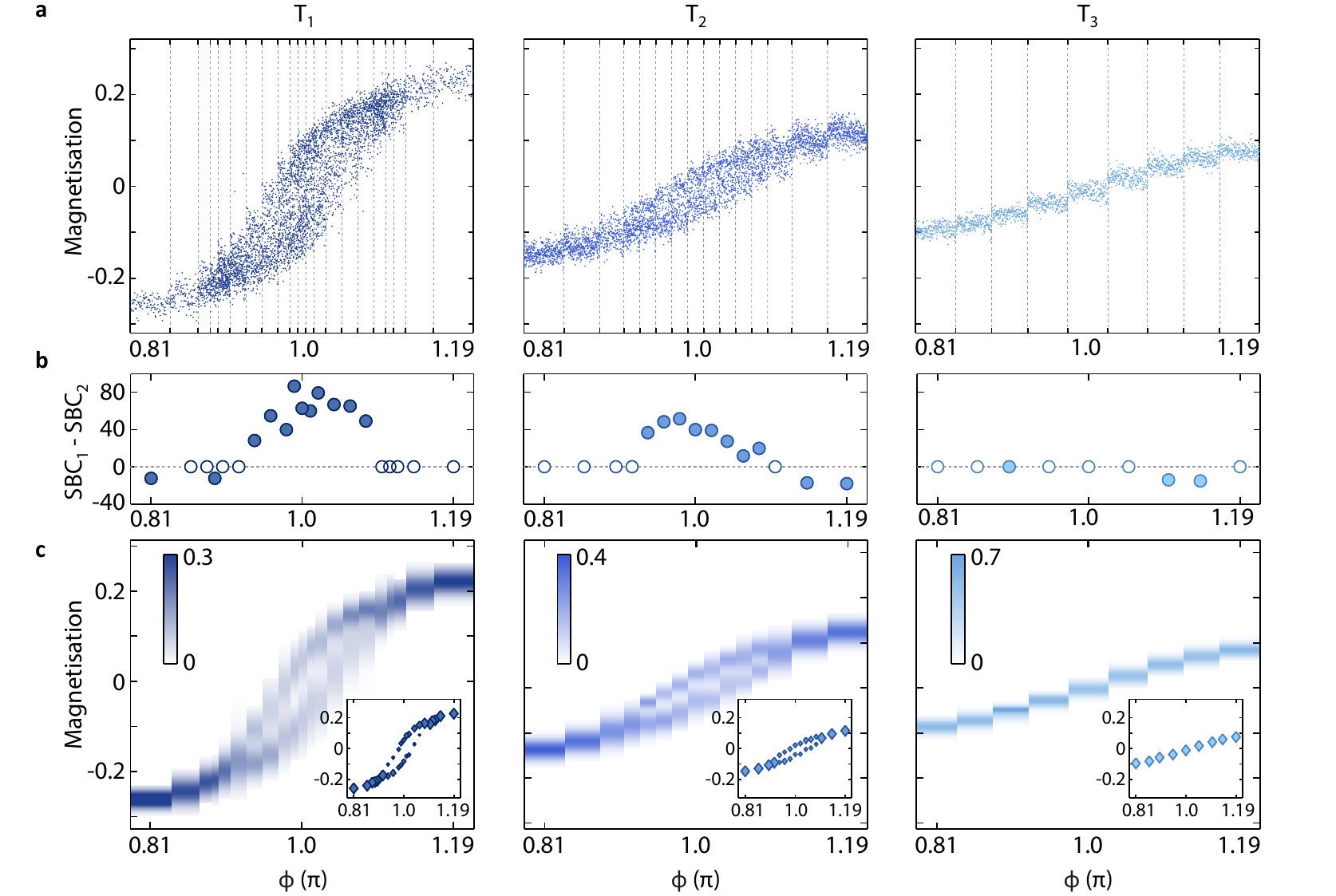}
\flushleft
\caption{\textbf{Raw data, information criteria and fit results of the statistical distributions.} \textbf{a}, Single shot measurements of the magnetisation as a function of the flux strength for the three different initial temperatures. \textbf{b}, Differences of Schwarz-Bayes criteria for a uni- and bimodal Gaussian probability distribution are plotted for each given flux value. As indicated by empty circles, for some cases the bimodal fit fails to converge and a unimodal fit has to be assumed as the best model. \textbf{c}, Resulting probability fits showing good agreement with the histograms from Fig.\,\ref{Figure3}\textcolor{darkblue}{b}. The extracted maxima of the probability distribution (see Fig.\,\ref{Figure3}\textcolor{darkblue}{c}) are shown in the insets.}
\label{ClusteringCriteria}
\end{figure*}
\begin{figure}
\hypertarget{fig:s8}{}
\centering
\includegraphics{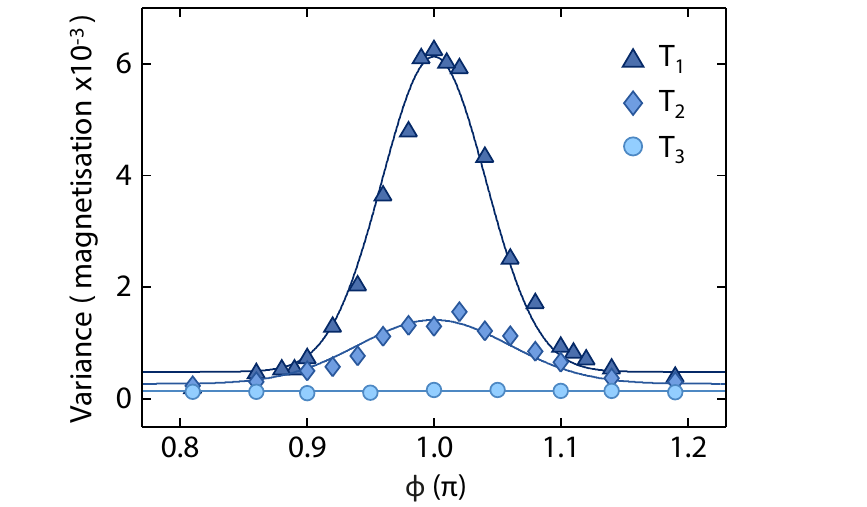}
\flushleft
\caption{\textbf{Variances.} Fluctuation of the magnetisation for the three temperatures demonstrating the symmetry breaking for fluxes close to $\Phi\,{=}\,\pi$. Each data point corresponds to the variance of the 1D-magnetisation data for the respective flux value. The solid lines are Gaussian fits to the data.}
\label{Variance}
\end{figure}
As described in the main text, the spontaneous breaking of the $\mathbb{Z}_2$ symmetry manifests itself in characteristic shot-to-shot fluctuations of the measured total magnetisation of the system. Therefore, a statistical analysis of the data is essential in order to extract reliable information about properties of the raw data distribution plotted in Fig.\,\hyperlink{fig:s7}{S7a}.

For this purpose, we fit a one-dimensional Gaussian probability distribution with $k_M \,{=}\, 1, 2$ modes to the magnetisation data for each flux value. With such a soft clustering method, the actual number and properties of modes  in the 1D-distributions can be determined by comparing the Schwarz-Bayes criterion (SBC) for the respective fits [\hyperlink{link:SuppBib}{S3}]. Fig.\,\ref{ClusteringExample} illustrates a uni- and bimodal Gaussian distribution for the case of flux $\Phi\,{=}\,\pi$ for the measurement with lowest temperature, where $\text{SBC}_1 > \text{SBC}_2$ favors the bimodal model. In order to assure reliable results, each fit is replicated ten times with random starting parameters, selecting the most likely output. Furthermore, the obtained parameters are averaged ten times so that deviations due to the randomness of the initial fitting parameters can be ruled out.
In Fig.\,\hyperlink{fig:s7}{S7b} the differences of the SBC for uni- and bimodal fits are plotted. For cases where $\text{SBC}_1 \le \text{SBC}_2$ ($\text{SBC}_1 > \text{SBC}_2$) a unimodal (bimodal) model is favored. Note that multimodal distributions with $k_M\,{\ge}\,3$ are not considered here since they have proven to be always less favorable as compared to the cases $k_M \,{=}\, 1$ and $k_M \,{=}\, 2$.

Spontaneous symmetry breaking for fluxes close to $\pi$ is clearly indicated for the temperatures $T_1$ and $T_2$ by favoring bimodal probability distributions. On the contrary, no symmetry breaking can be observed for the temperature $T_3$ as the unimodal fit is always favored (in spite of the outlier for $\Phi\,{=}\,0.9\,\pi$, where the SBC are nearly equal and the two resulting Gaussian modes strongly differ in width and amplitude, thus hinting at a remaining discrepancy in the evaluation of the respective fits). This is a decisive evidence of the phase transition from an ordered, ferromagnetic to an unordered, paramagnetic state. The resulting Gaussian density distributions (Fig.\,\hyperlink{fig:s7}{S7c}) are in good agreement with the statistical representation of the data in Fig.\,\hyperlink{fig:Figure3}{3b}. In Fig.\,\hyperlink{fig:Figure3}{3c} (and again in the inset of Fig.\,\hyperlink{fig:s7}{S7c})  the maxima of the obtained Gaussian distributions are plotted. In the case of bimodal distributions, the point size represents the ratio of amplitudes of the respective Gaussian, emphasising the smaller population of the metastable minimum for the measurement at $T_1$.

Another indication for the disappearance of spontaneous symmetry breaking for larger temperatures is the behaviour of the variance of the magnetisation measurements as shown in Fig.\,\ref{Variance}. Here, a notable rise of fluctuations for fluxes close to $\pi$ is evident for $T_1$, and, although less distinct, for $T_2$, while the fluctuations remain constantly small for $T_3$.

\section{Free Energy}
In this section we discuss the thermodynamic behaviour of the frustrated lattice system. We use a weak-coupling approximation of the free energy, which contains the non-interacting contribution and the first-order term, as discussed in Ref. [\hyperlink{link:SuppBib}{S4}].

\subsection{Symmetric case}

We consider the three-dimensional lattice dispersion in the fully frustrated case, for $\Phi=\pi$ and $|J|\,{=}\,|J'|$:
\begin{equation}
\begin{split}
\varepsilon({\bf k})\,{=}\, +&2|J|\cos\big(d k_{y}\big)\vphantom{\sqrt{3}k_{x}}\\
                            +&2|J|\cos\big(d [\sqrt{3}k_{x} - k_{y}]/2\big)\\
                            +&2|J|\cos\big(d [\sqrt{3}k_{x} + k_{y}]/2\big)+k_{z}^{2}/(2m).
\end{split}
\end{equation}
In order to map the system on a weakly interacting 3D Bose system, we expand the dispersion around the two minima ${\bf k}_{0,\texttt{A}/\texttt{B}}$ to second order, and write the curvature as an effective mass
\begin{equation}
\tilde{\varepsilon} ({\bf \xi}_{\texttt{i}}) = \frac{\xi_{\texttt{i},x}^{2}}{2 m_{x}}+\frac{\xi_{\texttt{i},y}^{2}}{2 m_{y}} +\frac{\xi_{\texttt{i},z}^{2}}{2 m_{z}},
\end{equation}
where ${\bf \xi}_{\texttt{i}} \equiv {\bf k} - {\bf k}_{0,\texttt{i}}$, and $\texttt{i}\,{=}\,\texttt{A},\texttt{B}$ with \texttt{A} and \texttt{B} denoting the two distinct minima in the dispersion. The in-plane masses are $m_{x}\,{=}\,m_{y}\,{=}\,m_{J}\,{=}\,2\hbar^{2}/(3d^{2}|J|)$, while the mass along the tube is simply the bare mass, $m_{z}\,{=}\,m$. The effective 3D density is related to the 1D density in the tubes by $n_{_{\text{3D}}}\,{=}\,n_{_{\text{1D}}}2/(\sqrt{3}d^{2})$. In analogy to the isotropic 3D case, we define the two thermal wavelengths
\begin{eqnarray}
\lambda_{m} &=& h/\sqrt{2\pi m k_{B} T}\\
\lambda_{J} &=& h/\sqrt{2\pi m_{J} k_{B} T}
\end{eqnarray}
and $\lambdabar\,{=}\,(\lambda_{m} \lambda_{J}^{2})^{1/3}$. We now consider a thermal distribution of non-interacting bosons. In analogy to the regular Bose gas we find for the density of excited states in the minima \texttt{A} and \texttt{B}
\begin{figure}
\hypertarget{fig:s9}{}
\includegraphics{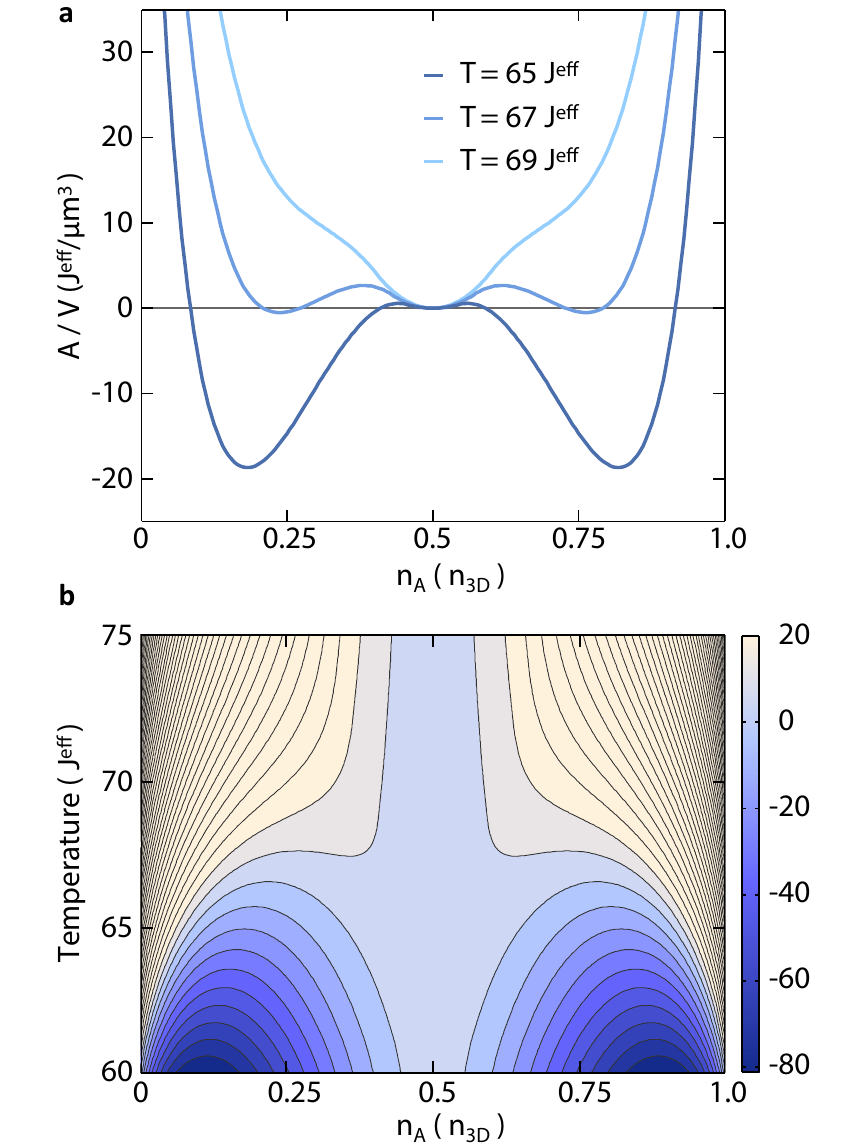}
\caption{\textbf{Free Energy in dependence of the density imbalance.} \textbf{a}, Free energy $A/V\,{=}\,(A_{0} + A_{1})/V$ per volume for  the parameters of the experiment, given in the text, as a function of the density $n_{\texttt{A}}$ of particles near minimum \texttt{A} of the dispersion for the temperatures $T_{1} =65\,J^{\text{eff}}$, $T_{2}\,{=}\,67\,J^{\text{eff}}$ and $T_{3}\,{=}\,69\,J^{\text{eff}}$. We keep the total density $n_{_{\text{3D}}}\,{=}\,n_{\texttt{A}} + n_{\texttt{B}}$ fixed and the energy $A(n_{\texttt{A}}\,{=}\,n_{_{\text{3D}}}/2, T)$ is set to zero. \textbf{b}, Contour plot of the free energy $A/V$ (in units of $J^{\text{eff}}/\mu\text{m}^3$) for a wider temperature range. The two new minima appear symmetrically around $n_{\texttt{A}}\,{=}\,n_{_{\text{3D}}}/2$, as the temperature is lowered.
\label{FE}}
\end{figure}
\begin{eqnarray}
n_{e,\texttt{i}} = \frac{1}{\lambdabar^{3}} g_{3/2}(\textsc{z}_{\texttt{i}})
\end{eqnarray}
with the Bose function $g_p(\textsc{z})\,{=}\,\sum_{l=1}^\infty\textsc{z}^l/l^p$ and the corresponding fugacities $\textsc{z}_{\texttt{i}}\,{=}\,\exp(\mu_{\texttt{i}}/k_B T)$. The chemical potentials $\mu_{\texttt{i}}$ control the densities $n_{\texttt{i}}$ in each minimum. The free energy of the non-interacting system is $A_{0}\,{=}\,A_{0,\texttt{A}}+A_{0,\texttt{B}}$, where
\begin{equation}
\frac{A_{0,\texttt{i}}}{V}  \,{=}\,
\begin{cases}
- \frac{k_{B} T}{\lambdabar^{3}} g_{5/2}(\textsc{z}_{\texttt{i}}) + n_{\texttt{i}} k_{B} T \ln \textsc{z}_{\texttt{i}} & \text{if } \textsc{z}_{\texttt{i}}<1\vphantom{\bigg[\bigg]}\\
- \frac{k_{B} T}{\lambdabar^{3}} g_{5/2}(1) & \text{if } \textsc{z}_{\texttt{i}}=1\nonumber
\end{cases}
\end{equation}
In order to account for the interaction, we include the first-order term in the effective 3D interaction strength $g_{_{\text{3D}}}$, which is related to the 1D interaction $g_{_{\text{1D}}}$ in the tubes by $g_{_{\text{3D}}}\,{=}\,g_{_{\text{1D}}}2d^{2}/\sqrt{3}$. As discussed in Ref.\,[\hyperlink{link:SuppBib}{S4}], the first order correction to $A/V$ is
\begin{eqnarray}
\frac{A_{1}}{V} &=& \frac{g_{_{\text{3D}}}}{2}\big[2(n_{\texttt{A}} + n_{\texttt{B}})^{2} - n_{0,\texttt{A}}^{2} - n_{0,\texttt{B}}^{2}\big]
\end{eqnarray}
where $n_{0,\texttt{A}}$ and $n_{0,\texttt{B}}$ are the condensate densities in minimum \texttt{A} and \texttt{B}, respectively.
In Fig.\,\ref{FE}, we plot the free energy per volume $A/V\,{=}\,(A_{0}\,{+}\,A_{1})/V$, for $J\,{=}\,J^{\text{eff}}\,{=}\,k_{B}\,{\times}\,0.26\,\text{nK}$, and for a fixed total density of $n_{_{\text{3D}}}\,{=}\,17\,{\mu}\text{m}^{-3}$, corresponding to a 1D density of $n_{_{\text{1D}}}\,{=}\,6\,{\mu}\mathrm{m}^{-1}$. With a 1D interaction strength of $g_{_{\text{1D}}}\,{=}\,23.4\,J^{\text{eff}}\mu m$, this results in an effective 3D interaction strength of $g_{_{\text{3D}}}\,{=}\,6.4\,J^{\text{eff}}\mu m^3$. As the temperature $T$ is lowered, the free energy develops two minima symmetrically around $n_{1}\,{=}\,n_{_{\text{3D}}}/2$, indicating the onset of spontaneous breaking of a $\mathbb{Z}_{2}$ symmetry. Furthermore, we see that within this approximation the free energy barrier is of the order of $g_{_{\text{3D}}} n_{0} N_{0}$, where $n_{0}$ and $N_{0}$  denote the density and the number of condensed particles, respectively. When the condensate fraction approaches 1, the energy barrier per particle becomes $g_{_{\text{3D}}} n_{0}\,{\approx}\,g_{_{\text{3D}}} n_{_{\text{3D}}}$ which is of the order of $k_{B}\times 28\,\text{nK}$ or $108\,J^{\text{eff}}$. Since this energy is large compared to the temperature estimates of the experiment, it can protect the metastable states that are seen following the quench. We also note that in this estimate the breaking of the $\mathbb{Z}_{2}$ and the $U(1)$ symmetry occur at the same temperature, because it is the condensate fraction that is responsible for generating two minima in the free energy.

\subsection{Biased case}
We now consider the case where the minima of the dispersion relation are not degenerate, but have an energy difference of $\Delta \equiv \varepsilon({\bf k}_{0,\texttt{B}}) - \varepsilon({\bf k}_{0,\texttt{A}})$ resulting from a flux value different from $\Phi\,{=}\,\pi$. An approximate relation between tilt energy $\Delta$ and flux strength $\Phi$ is $\Delta\,{=}\,10.5\,J^{\text{eff}}\times (\Phi/\pi -1)$. We choose the energy minima of the dispersion such that $\varepsilon({\bf k}_{0,\texttt{A}})\,{=}\,|\Delta|$ and $\varepsilon({\bf k}_{0,\texttt{B}})=0$ for $\Delta<0$, and $\varepsilon({\bf k}_{0,\texttt{A}})\,{=}\,0$ and $\varepsilon({\bf k}_{0,\texttt{B}})\,{=}\,\Delta$ for $\Delta>0$.  Using the same approximation as in the previous section we find the following expression for the free energy
\begin{equation}
\begin{split}
A/V = - \vphantom{\Big(}\frac{k_BT}{\lambdabar^{3}} &\big[g_{5/2}(\textsc{z}_{\texttt{A}}) +g_{5/2}(\textsc{z}_{\texttt{B}})\big]\\
      + \vphantom{\Big(}k_{B}T                                                  &\big[n_{\texttt{A}}\ln(\textsc{z}_{\texttt{A}}) + n_{\texttt{B}} \ln(\textsc{z}_{\texttt{B}})\big]\\
      + \vphantom{\Big(}\frac{g}{2}                                             &\big[2(n_{\texttt{A}}+n_{\texttt{B}})^{2}-n_{0,\texttt{A}}^{2}-n_{0,\texttt{B}}^{2}\big].\label{FEbias}
\end{split}
\end{equation}
If only the density $n_{_{\text{3D}}}\,{=}\,n_{\texttt{A}} + n_{\texttt{B}}$ is given, as it is the case for the experiment, only one of the density fractions, $n_{0,\texttt{A}}$ or $n_{0,\texttt{B}}$, can be non-zero. As is apparent from equation \,(\ref{FEbias}), the system can always lower its energy by condensing all atoms into only one of the two minima.
\begin{figure}
\hypertarget{fig:s10}{}
\hypertarget{fig:FEbiasplot}{}
\includegraphics{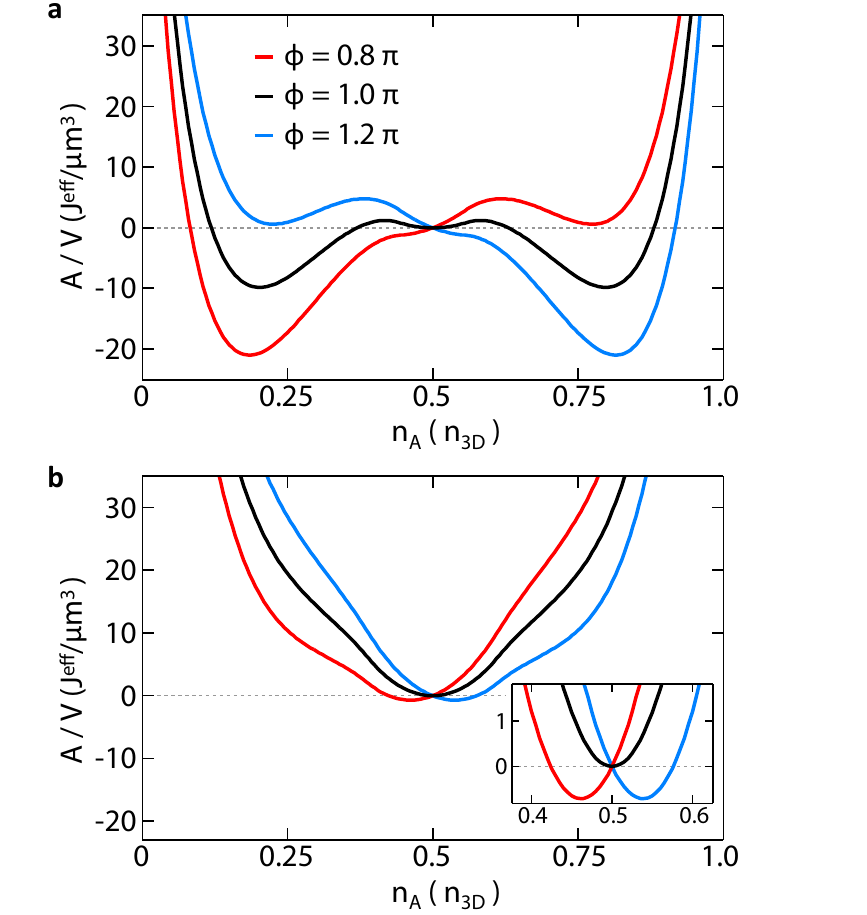}
\caption{\textbf{Free Energy behaviour in dependence of the flux.} Free energy $A/V$ per volume as a function of the density $n_{\texttt{A}}$ for various flux strengths $\Phi$ and a temperature of \textbf{a}, $T\,{=}\,66\,J^{\text{eff}}$ and \textbf{b}, $T\,{=}\,70\,J^{\text{eff}}$. We keep the total density $n_{_{\text{3D}}}\,{=}\,n_{\texttt{A}}\,{+}\,n_{\texttt{B}}$ fixed. The energy $A(n_{\texttt{A}}\,{=}\,n_{_{\text{3D}}}/2, \Delta)$ is set to zero. For the smaller temperature in \textbf{a}, a local minimum persists for finite tilt energy, indicating the two degenerate minima that exist for the symmetric case. For the higher temperature in \textbf{b} only one minimum can be seen, indicating that the system is supercritical.
\label{FEbiasplot}}
\end{figure}

\begin{figure}
\hypertarget{fig:s11}{}
\includegraphics{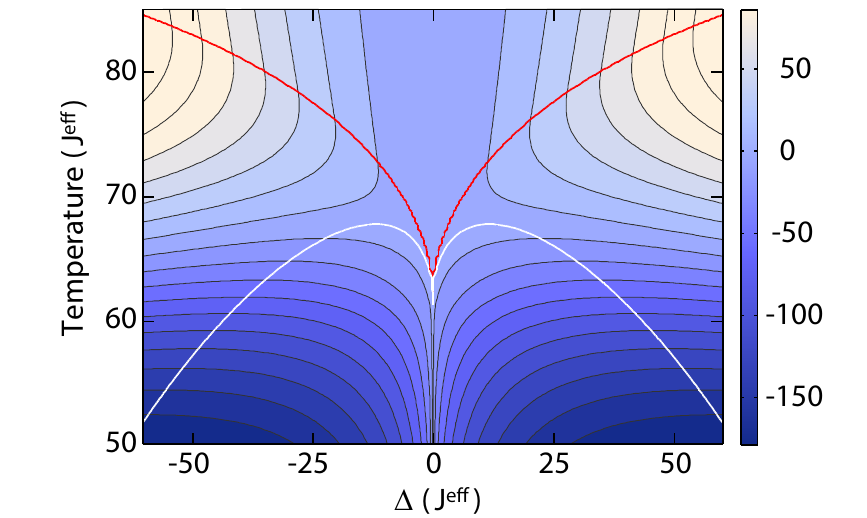}
\caption{\textbf{Free energy A/V as a function of temperature and tilt energy.} Here, only the total density $n_{_{\text{3D}}} \,{=}\,n_{\texttt{A}} + n_{\texttt{B}}$ is held constant. Above the critical point, tilting the dispersion increases both the free energy (plotted here in units of $J^{\text{eff}}{/}\mu\text{m}^3$), and the phase space density in the lower minimum, leading to condensation. While the phase transition is depicted by the red line, the white line marks the condition $g n_0 \,{=}\,\Delta$, corresponding to the metastability of the upper minimum. Below the critical point, the free energy decreases when the dispersion is tilted, indicating an instability.
\label{FElarge}}
\end{figure}

\begin{figure}
\hypertarget{fig:s12}{}
\includegraphics{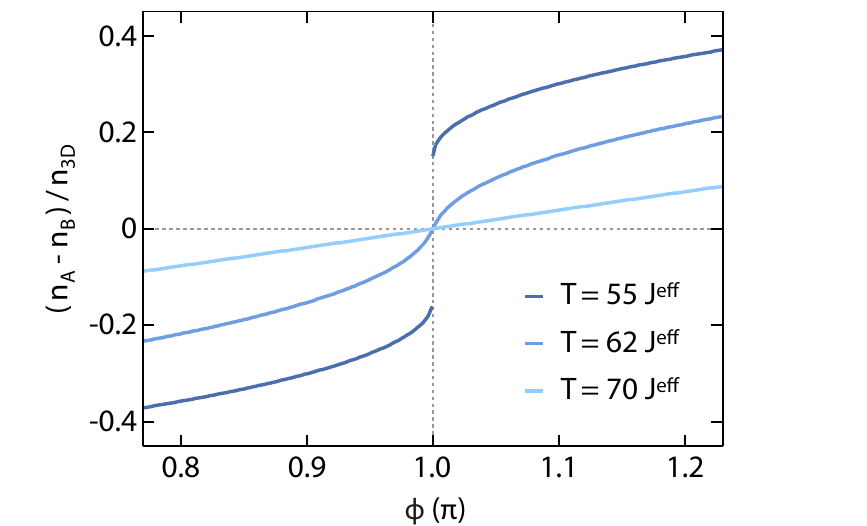}
\caption{\textbf{Density imbalance as a function of the flux}. The density imbalance $(n_{\texttt{A}} - n_{\texttt{B}})/n_{_{\text{3D}}}$ between the two minima in the dispersion is calculated for three different temperatures as a function of the flux strength $\Phi$. For subcritical temperatures of the $\mathbb{Z}_{2}$ transition, the density imbalance has a discontinuity at $\Phi=\pi$. For supercritical temperatures we see a linear dependency.
\label{densfrac}}
\end{figure}

We first hold the individual densities fixed to demonstrate the behaviour of the free energy described in the main text. In Fig.\,\ref{FEbiasplot} we show the free energy $A/V$ as a function of $n_{\texttt{A}}$, with $n_{\texttt{A}} + n_{\texttt{B}}$ held fixed, for different flux strengths $\Phi$. In Fig.\,\hyperlink{fig:FEbiasplot}{\ref{FEbiasplot}a}, a local minimum persists for a finite tilt. Here the temperature is low enough, that the system is condensed at finite tilt of the system away from $\Phi=\pi$. In Fig.\,\hyperlink{fig:FEbiasplot}{\ref{FEbiasplot}b} we choose a higher temperature, resulting in only one global minimum being present for any tilt. We now only hold the total density $n_{\texttt{A}} + n_{\texttt{B}}$ fixed so the fugacities are given by $\textsc{z}_{1} \,{=}\,\textsc{z}$ and $\textsc{z}_{2} \,{=}\,\textsc{z}\exp(-|\Delta|/k_{B} T)$, for $\Delta > 0$, and by $\textsc{z}_{1} \,{=}\,\textsc{z}\exp(-|\Delta|/k_{B} T)$ and $\textsc{z}_{2} \,{=}\,\textsc{z}$ for $\Delta <0$. The density of excited states is related to the fugacity $\textsc{z}$ through
\begin{equation}
n_{e,\texttt{A}} +n_{e,\texttt{B}} = \Big[g_{3/2}(\textsc{z}) + g_{3/2}\big(\textsc{z}\,\mathrm{e}^{-|\Delta|/{k_{B}T}}\big)\Big]/\lambdabar^{3}.
\end{equation}
The free energy is then given by
\begin{equation}
\begin{split}
A/V =  -\frac{k_{B}T}{\lambdabar^{3}}&\Big[ g_{5/2}(\textsc{z})+g_{5/2}\big(\textsc{z}\,\mathrm{e}^{-|\Delta|/{k_{B}T}}\big)\Big]\\
 + k_{B}T&(n_{\texttt{A}}+n_{\texttt{B}})   \ln \textsc{z}  - n_{\texttt{i}}\Delta\\
 + g&\big[(n_{\texttt{A}}+n_{\texttt{B}})^{2} - n_{0,\texttt{i}}^{2}/2]\label{FEbias2}
\end{split}
\end{equation}
where $\texttt{i}\,{=}\,\texttt{B}$ if $\Delta >0$, and $\texttt{i}=\texttt{A}$ if $\Delta <0$. In Fig.\,\ref{FElarge} we show the free energy for the same parameters as in the previous section, in dependence of the temperature $T$ and the tilt energy $\Delta$, while keeping only the total density $n_{_{\text{3D}}}$ fixed rather than the individual densities. Note that here we span a much larger parameter range of the tilt energy than is experimentally accessible (compare Fig.\,\ref{Figure4} in the main text). For each temperature, $A(\Delta=0, T)$ has been set to zero. If the system is above the critical point, the free energy increases as the tilt energy is varied away from $\Delta =0$. If the system is below the critical point, the free energy decreases, indicating an instability towards a $\mathbb{Z}_{2}$ symmetry broken state. The critical temperature increases when a non-zero tilt is chosen, as indicated by the red line.
When the dispersion is tilted, the phase space density increases in the lower minimum, which results in a higher condensation temperature. For the non-interacting system the ratio of the critical temperatures for large tilt and no tilt is $T_{c, \Delta\rightarrow \infty} / T_{c, \Delta \,{=}\,0} \,{=}\,2^{2/3}$. Furthermore, the condition $|\Delta| \,{=}\,g n_{0}$ is shown by a white line. As discussed, this gives the order of magnitude of $\Delta$ for which the higher minimum of the free energy vanishes. This estimate is accurate for small temperatures, and gives an approximate energy scale for higher temperatures. We see that it is a large scale compared to the tilt energies studied in experiment. Therefore, the higher minimum is typically stable, once the system is subcritical. In Fig.\,\ref{densfrac} we show the density fraction $(n_{\texttt{A}} - n_{\texttt{B}})/n_{_{\text{3D}}}$ as function of the flux strength $\Phi$ for different temperatures. For high temperatures, a linear response to the tilt can be seen, while the appearance of a discontinuity for subcritical temperatures is indicative of a phase transition.

\begin{figure}
\hypertarget{fig:s13}{}
\includegraphics{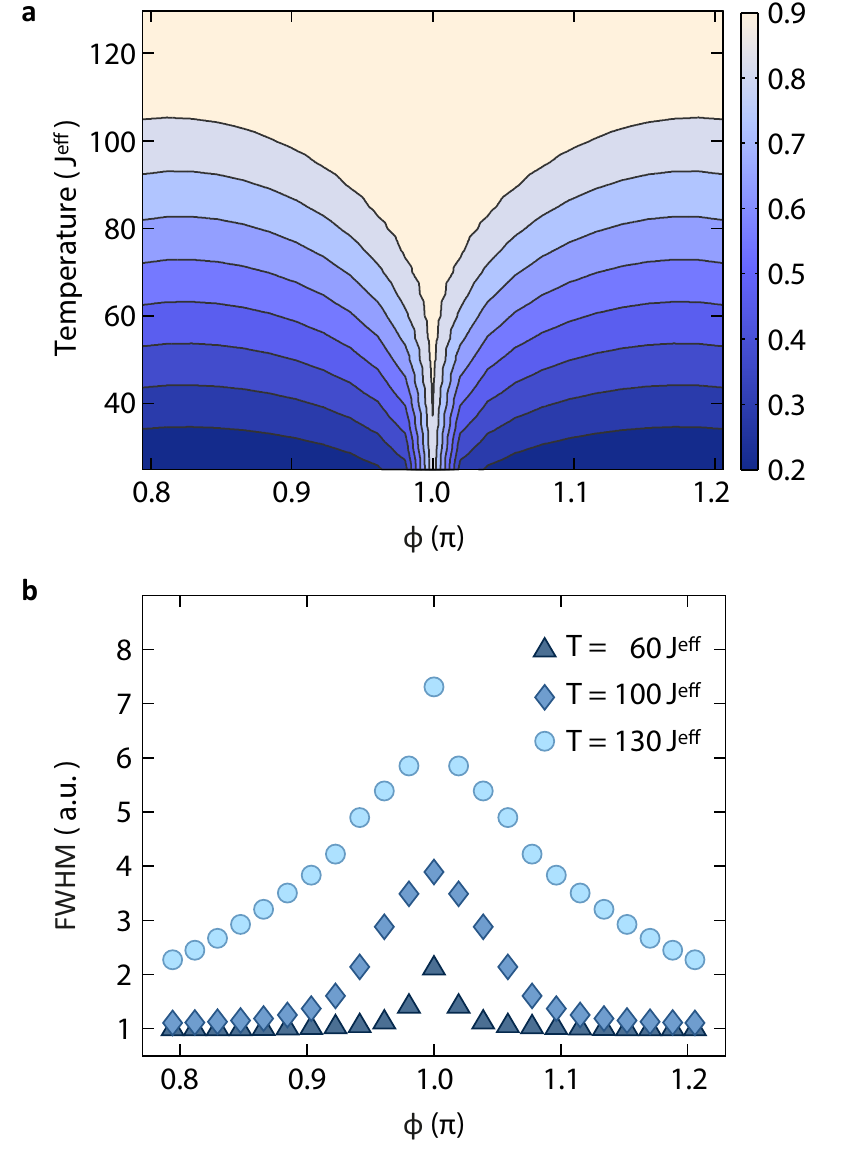}
\caption{\textbf{Calculations for the exact band structure in a non-interacting approximation.} Data for $N_{\mathrm{Tot}}/N_{\mathrm{Sites}}=90$ and $N_{\mathrm{Sites}}\,{=}\,17\times 17$. \textbf{a,} The density of excited atoms $n_{\mathrm{exc}}\equiv N_{\mathrm{exc}}/N_{\mathrm{Tot}}$ is strongly enhanced at a flux strength of $\pi$, resulting in a cusp. \textbf{b,} The FWHM of the momentum distribution increases close to $\Phi\,{=}\,\pi$ and at higher temperatures, pointing at a decrease of $U(1)$ long-range order. The values are normalised to the lowest result.}
\label{fig:Nexc_exactbandstructure}
\end{figure}

\begin{figure}
\hypertarget{fig:s14}{}
\includegraphics{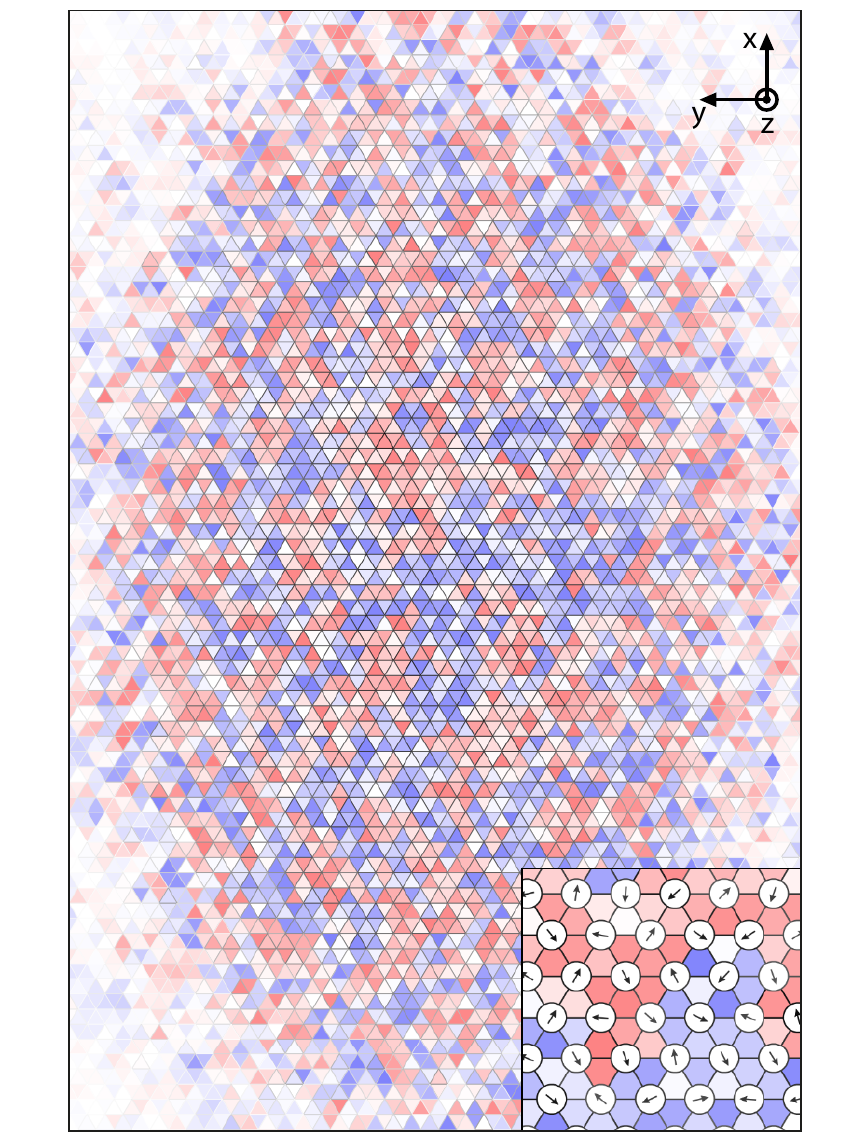}
\caption{\textbf{Monte-Carlo sample snapshot.} Sample snapshot of thermal equilibrium ensemble for flux strength $\Phi\,{=}\,\pi$ at $T\,{=}\,37\,J^{\text{eff}}$ (so $T\,{>}\,T_{\text{Ising}})$, where the $xy$-plane is cutting the tubes at the center of the trap: triangular plaquettes with negative (positive) bosonic currents in red (blue). Higher (lower) color intensity represents higher (lower) absolute values of the bosonic current. In addition, regions with lower density are covered in white haze. Inset: Sample from center with arrows representing the phase at each lattice site.}
\label{fig:monte-carlo-snapshot}
\end{figure}

\section{\label{sec:Tcrit}Estimation of the critical temperature}

One can obtain a reasonable estimate for the number of Bose-condensed atoms at higher temperatures by approximating the system as non-interacting and neglecting the trap in the $xy$-plane. In that case, the total number of atoms is given by the Bose statistics for the dispersion relation $\varepsilon(\mathbf{k})$ together with the discrete level structure of the harmonic trap in $z$ direction,
\begin{equation}
	N_{\mathrm{Tot}}=\sum_{k_x,k_y}\sum_{n_z=0}^{\infty} \Big( \mathrm{e}^{    \frac{\varepsilon(k_x,k_y)+\hbar\omega_z(n_z+1/2)-\mu}{k_{\mathrm{B}} T}   }-1\Big)^{-1} \,.
\end{equation}
The Bose--condensate is, for given chemical potential $\mu$, identified with the atom number $N_0$ in the minimum of the dispersion relation. To work at fixed particle number, we adjust $\mu$ until the density equals a desired value, for which we use a realistic number ($N_{\mathrm{Tot}}/N_{\mathrm{Sites}}=90$ in a rhombic lattice of $N_{\mathrm{Sites}}=17\times 17$).
Further, we assume the experimental value $\hbar\omega_z/J^{\text{bare}}=3$, and we use the effective tunnelings for a given shaking modulation $\delta$ (see Fig.\,\hyperlink{fig:s1}{S1}). We normalise all quantities to the magnitude of the resulting tunneling matrix elements $J^{\text{eff}}$. The results for the number of excited atoms in the exact band structure are shown in Fig.\,\hyperlink{fig:s13}{\ref{fig:Nexc_exactbandstructure}a}. The condensate is much stronger depleted at fluxes close to $\pi$, indicating a stronger loss of $U(1)$ long-range order and a lower critical point for Bose condensation. This results in the same cusp-like behaviour as obtained with the calculations in the weakly-interacting system with a harmonic approximation to the band structure (Fig.\,\hyperlink{fig:Figure5}{\ref{Figure5}b} of the main text).

From the occupation at different $\mathbf{k}$ modes, one can also compute the peak width of the momentum distribution, similar to what is extracted from the experimental time-of-flight images and plotted in Fig.\,\hyperlink{fig:Figure5}{\ref{Figure5}a} of the main text. The result is shown in Fig.\,\hyperlink{fig:s13}{\ref{fig:Nexc_exactbandstructure}b}. It reproduces well the qualitative behaviour of the experimental findings: the momentum peak gets broader closer to $\pi$ flux and at higher temperatures, pointing at a decrease of the $U(1)$ long-range order.

\section{Monte-Carlo Simulation}

In order to study the equilibrium states of the three dimensional ultracold ensemble, we simulate the system using classical Monte-Carlo (Metropolis algorithm). We start with a system of $42\times75$ tubes, where each tube is discretised into $33$ sites using a discretisation length of $a_{z}=1\,\mu\text{m}$. This introduces an additional effective tunneling term in z-direction $J_{z}=\hbar^{2}/(2ma_{z}^{2})=10.7\,J^{\text{eff}}$. Furthermore, the on-site repulsive interaction constant has to be rescaled to $U_{\text{site}}=g_{_{\mathrm{1D}}}/a_{z}$.

The system is initialised using one of the two-fold degenerate classical ground states at zero flux with an initial total number of atoms of $2\times10^{5}$. To ensure that the total number of atoms remains constant, we set the chemical potential to $\mu=230\,J^{\text{eff}}$. We then perform single-site updates, where sites are chosen randomly and changes to the real and imaginary parts of the wavefunction are generated by sampling from a normal distribution whose width is adjusted to approach a step acceptance rate of roughly one half. After a thermalisation process that, depending on the temperature, consists in $10^{5}-10^{6}$ Monte-Carlo steps per site (MCS), we start taking $100$ snapshots of the systems. Between subsequent samples we perform sufficiently many MCS for both samples being completely uncorrelated: correlation is eradicated in under ten MCS, whereas our sampling frequency is between 100 and 1000 MCS. For each sample we compute the chirality which is given by the relative visibility of the integrals over both interference peaks in quasimomentum space using a triangular mask. We repeat this process for flux values $\in\left[0.81\pi,1.19\pi\right]$ and obtain the chirality as a function of the flux. The results are shown in Fig.\,\ref{Figure4}.

\renewcommand{\bibnumfmt}[1]{[S{#1}]}

\end{document}